\documentclass[article,twocolumn,apl,superscriptaddress,amsmath,amssymb,floatfix]{revtex4-2}

\usepackage{textcomp}
\usepackage{amsmath}
\usepackage{physics}
\usepackage{amsfonts}
\usepackage{amssymb}
\usepackage{bm}
\usepackage{MnSymbol}

\usepackage{xspace}
\usepackage{graphicx} 
\usepackage[dvipsnames]{xcolor}
\usepackage[separate-uncertainty=true,multi-part-units=single,retain-unity-mantissa=false]{siunitx}
\usepackage[version=4]{mhchem}
\usepackage[colorlinks, citecolor={blue}, linkcolor={black}, urlcolor={blue}]{hyperref}
\usepackage{xfp} 
\newcommand\ExtendedData{
    \xdef\preextfigures{\arabic{figure}}
    \renewcommand\thefigure{\textbf{\fpeval{\arabic{figure}-\preextfigures}}}
    \renewcommand{\figurename}{\textbf{\extlbl}}
}
\newcommand\SupplementaryMaterials{
    \xdef\presupfigures{\arabic{figure}}
    \xdef\presupsections{\arabic{subsection}}
    \renewcommand\thefigure{\textbf{\fpeval{\arabic{figure}-\presupfigures}}}
    \renewcommand\thesubsection{\fpeval{\arabic{subsection}-\presupsections}}
    \renewcommand{\figurename}{\textbf{\supplbl}}
}

\newcommand{\figlbl}{Fig.}
\newcommand{\extlbl}{Extended~Data~Fig.} 
\newcommand{\supplbl}{Supplementary~Fig.}

\renewcommand{\figurename}{\textbf{\figlbl}} 
\renewcommand{\thefigure}{\textbf{\arabic{figure}}}
\newcommand{\figtitle}[1]{\textbf{#1}\xspace} 

\newcommand{\panel}[1]{\textbf{#1}\xspace}
\newcommand{\moire}{moir{\'e}\xspace}
\newcommand{\capmoire}{Moir{\'e}\xspace}

\begin{document}

\title{Superconductivity and strong interactions in a tunable \moire quasiperiodic crystal}

\author{Aviram Uri}
\thanks{These authors contributed equally.}
\author{Sergio C. de la Barrera}
\thanks{These authors contributed equally.}
\author{Mallika T. Randeria}
\thanks{These authors contributed equally.}
\author{Daniel Rodan-Legrain}
\thanks{These authors contributed equally.}
\author{Trithep Devakul}
\author{Philip J. D. Crowley}
\author{Nisarga Paul}
\affiliation{Department of Physics, Massachusetts Institute of Technology, Cambridge, MA, USA}
\author{Kenji Watanabe}
\affiliation{Research Center for Functional Materials, National Institute for Materials Science, 1-1 Namiki, Tsukuba 305-0044, Japan}
\author{Takashi Taniguchi}
\affiliation{International Center for Materials Nanoarchitectonics, National Institute for Materials Science, 1-1 Namiki, Tsukuba 305-0044, Japan}
\author{Ron Lifshitz}
\affiliation{Raymond and Beverly Sackler School of Physics \& Astronomy, Tel Aviv University, Tel Aviv 69978, Israel}
\author{Liang Fu}
\author{Raymond C. Ashoori}
\author{Pablo Jarillo-Herrero}
\affiliation{Department of Physics, Massachusetts Institute of Technology, Cambridge, MA, USA}

\begin{abstract}
Electronic states in quasiperiodic crystals generally preclude a Bloch description \cite{lesser2022emergence}, rendering them simultaneously fascinating and enigmatic.
Owing to their complexity and relative scarcity, quasiperiodic crystals are underexplored relative to periodic and amorphous structures.
Here, we introduce a new type of highly tunable quasiperiodic crystal easily assembled from periodic components.
By twisting three layers of graphene with two different twist angles, we form two \moire patterns with incommensurate \moire unit cells.
In contrast to many common quasiperiodic structures that are defined on the atomic scale \cite{janssen2018aperiodic, tsai2008icosahedral, steurer2004twenty}, the quasiperiodicity in our system is defined on \moire length scales of several nanometers.
This novel ``\moire quasiperiodic crystal'' allows us to tune the chemical potential and thus the electronic system between a periodic-like regime at low energies and a strongly quasiperiodic regime at higher energies, the latter hosting a large density of weakly dispersing states.
Interestingly, in the quasiperiodic regime we observe superconductivity near a flavor-symmetry-breaking phase transition \cite{sharpe2019emergent,wong2020cascade,zondiner2020cascade,zhou2021halfandquarter,chen2020tunable}, the latter indicative of the important role electronic interactions play in that regime. 
The prevalence of interacting phenomena in future systems with \textit{in situ} tunability is not only useful for the study of quasiperiodic systems, but it may also provide insights into electronic ordering in related periodic \moire crystals \cite{cao2018unconventional,andrei2021marvels,wu2018theory,wu2019identification,cea2021coulomb,khalaf2021charged,lewandowski2021pairing,chou2021correlation,lake2022pairing}.
We anticipate that extending this new platform to engineer quasiperiodic crystals by varying the number of layers and twist angles, and by using different two-dimensional components, will lead to a new family of quantum materials to investigate the properties of strongly interacting quasiperiodic crystals.

\end{abstract}

\maketitle

Quasicrystals are ordered solids that lack periodicity \cite{lifshitz2011symmetry}.
They may possess symmetries that are forbidden in periodic crystals \cite{janssen2018aperiodic, steurer2004twenty, tsai2008icosahedral}, in addition to those that are allowed in periodic crystals \cite{lifshitz2002square,lifshitz2003quasicrystals,koshino2022topological}.
In both cases, Bloch's theorem is generally inapplicable \cite{lesser2022emergence}, presenting significant challenges for understanding electronic correlations and topology in quasiperiodic systems.
Progress in these directions has led to predictions of new topological \cite{kraus2013fourdimensional,tran2015topological,huang2018quantum,spurrier2018semiclassical,else2021quantum,koshino2022topological} and superconducting \cite{sakai2017superconductivity} phenomena in quasicrystals.
However, experimental evidence of similar quantum phenomena in quasicrystals has been demonstrated only in a handful of cases, with superconductivity \cite{kamiya2018discovery}, topology \cite{dareau2017revealing,lohse2018exploring}, and quantum critical magnetic behavior \cite{deguchi2012quantum} being notable examples.
Moreover, the controlled synthesis of quasicrystals presents additional challenges.
A flexible material system for engineering quasiperiodicity can therefore facilitate rapid experimental advances and inspire new theoretical ideas.

\begin{figure*}
    \centering
    \includegraphics[width=180mm]{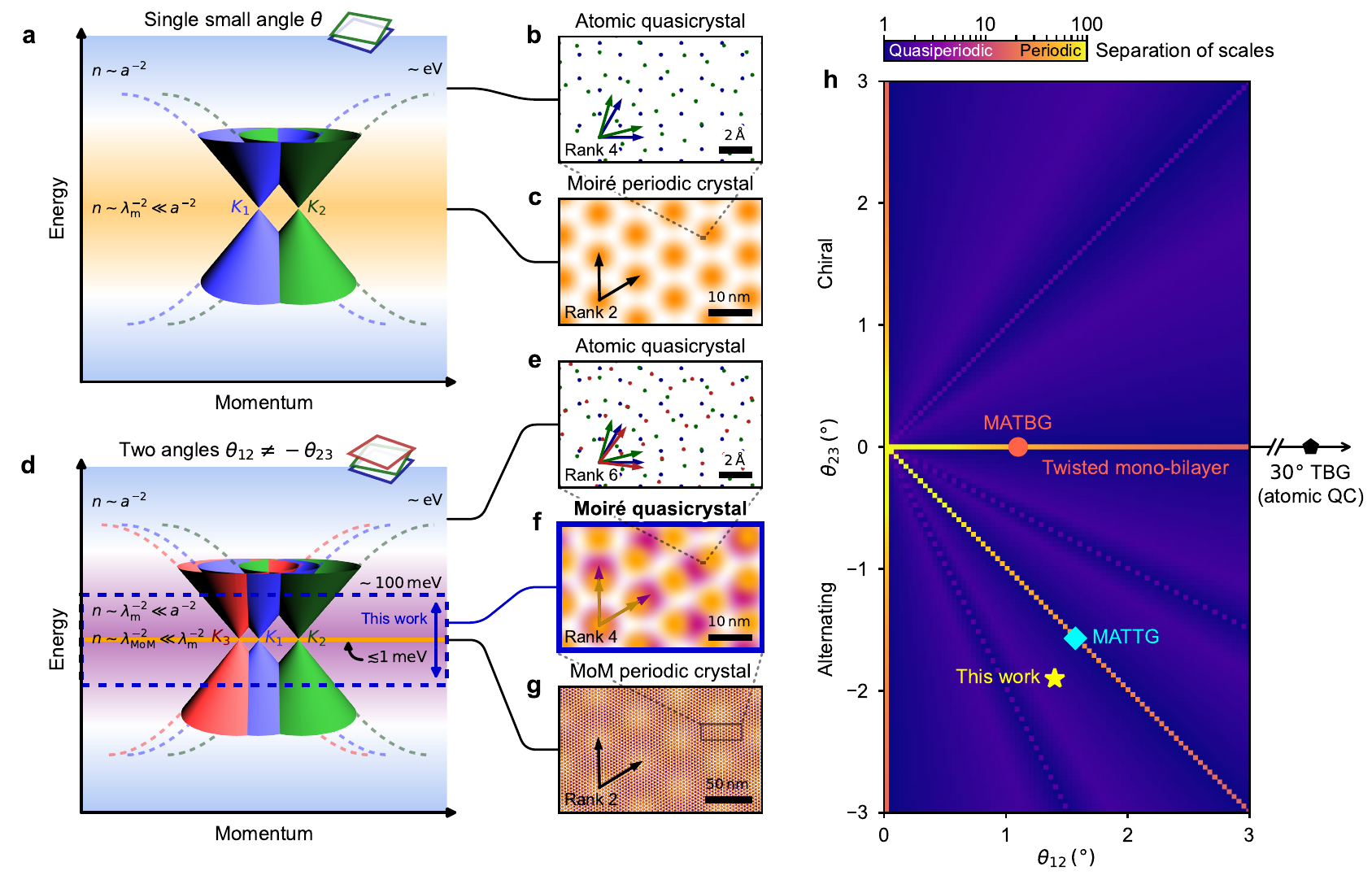}
    \caption{\figtitle{\capmoire quasiperiodicity.}
    \panel{a} Energy-momentum illustration of Dirac systems with a single, small twist angle. 
    The incommensurate atomic lattices form a rank-4 atomic quasicrystal, where the rank is the minimal number of primitive vectors required to describe the lattice (real space illustration in \panel{b}).
    The atomic details are relevant for the electronic structure only at high energies and carrier densities, $n$ (blue shading).
    At low densities (orange shading), the electronic structure is determined by an emergent rank-2 \moire periodic crystal (real space in \panel{c}).
    \panel{d} Same as \panel{a} for systems with two different twist angles. 
    At high carrier densities (blue shading) a rank-6 atomic quasicrystal is relevant (real space in \panel{e}).
    At intermediate densities (purple shading), the effective system comprises two pairwise \moire structures of similar length scales, forming a rank-4 \moire quasicrystal (real space in \panel{f}) -- the focus of this work.
    At ultra-low densities (thin orange line), a rank-2 ``\moire-of-\moire'' (MoM) periodic crystal may emerge (real space in \panel{g}), not our focus here.
    \panel{b-c,e-g} Colors indicate effective lattices, with arrows as lattice vectors. 
    \panel{c,f,g} Pink (orange) and white correspond to AA sites of layers 12 (23) and AB/BA arrangements, respectively.
    \panel{h} Separation of length scales, $\gamma(\theta_{12},\theta_{23})$, of TTG at intermediate densities, $n \sim \lambda_\text{m}^{-2}$.
    $\gamma \gg 1$ describes \moire periodic crystals (\panel{c}), whereas $\gamma \gtrsim 1$ results from two competing \moire lattices of similar length scales and describes MQCs (\panel{f}).
    }
    \label{fig:sos}
\end{figure*}

Layered assembly of van der Waals materials provides a convenient platform for lattice engineering that avoids the complications of conventional synthesis \cite{geim2013vanderwaals,andrei2021marvels}. 
One important example is twisted bilayer graphene (TBG) -- two layers of graphene twisted by a small angle (Fig.~\ref{fig:sos}\panel{a}).
TBG possesses a quasiperiodic atomic structure (Fig.~\ref{fig:sos}\panel{b}), but the low-energy electronic behavior is instead driven by an emergent long wavelength \moire periodicity \cite{bistrizter2011moire} (Fig.~\ref{fig:sos}\panel{c}).
Using three twisted layers of graphene, however, we can take advantage of the \moire length scale to generate a different type of quasiperiodicity that dominates the electronic behavior at relevant energies.
Specifically, three layers of graphene with two unequal twist angles (Fig.~\ref{fig:sos}\panel{d}) produce atomic-scale quasiperiodicity (Fig.~\ref{fig:sos}\panel{e}), but in contrast to TBG, the low energy electronic structure is determined by two \moire lattices that emerge from adjacent layers, rather than just one \cite{oka2021fractal,mao2021quasiperiodicity,cea2020bandstructure,shi2021moire,meng2022commensurate}.
Importantly, the two \moire lattices are generally incommensurate, leading to a qualitatively different quasiperiodic system (Fig.~\ref{fig:sos}\panel{f}).
We term this new class of incommensurate structures ``\moire quasiperiodic crystals'' (MQCs) or ``\moire quasicrystals'' for short.
\capmoire quasicrystals arise not from quasiperiodicity of atoms (Fig.~\ref{fig:sos}\panel{e}), but from incommensurability between more than one \moire lattice (Fig.~\ref{fig:sos}\panel{f}).
Crucially, \moire quasiperiodicity can be engineered by selecting twist angles and constituent materials.

We stress that here we do not focus on ``\moire of \moire'' or ``super\moire'' periodicity (Fig.~\ref{fig:sos}\panel{g}), an approximate long-wavelength periodicity emerging from two \moire lattices \cite{wang2019newgeneration,wang2019composite,zhu2020twistedtrilayer,zhang2021correlated,li2022symmetrybreaking}, which may be relevant at ultra-low energies (potentially below the disorder limit) and for larger twist angles.

Although many studies of twisted trilayer graphene (TTG) have focused on \moire periodic crystals, such as magic-angle TTG \cite{khalaf2019magic,park2021tunable,hao2021electricfield,cao2021pauli,kim2022evidence,liu2022isospin}, the majority of the space spanned by the two twist angles of TTG hosts MQCs.
Figure~\ref{fig:sos}\panel{h} shows this graphically based on a comparison of length scales (Methods~\ref{ssec:SOS}).
We define the separation of scales, $\gamma(\theta_{12},\theta_{23})$, as the minimal ratio (keeping $\gamma>1$) between the relevant length scales in TTG, $\lbrace\lambda_{12},\lambda_{23},\lambda_{13},a\rbrace$, where $\lambda_{ij} \approx a/\theta_{ij}$ are the \moire lattice constants, $\theta_{ij}$ is the angle between layers $i$ and $j$, and $a\approx\SI{0.246}{nm}$ is the graphene lattice constant.
In systems with a single small twist angle, $\gamma = \lambda_\text{m}/a \gg 1$, indicative of \moire periodic crystals (here, $\lambda_\text{m}$ is the \moire length).
In contrast, for $\theta_{12}\neq -\theta_{23}$ two competing \moire periodicities with comparable length scales give $\gamma \gtrsim 1$, signifying MQCs.

\begin{figure*}
    \centering
    \includegraphics[width=6.75in]{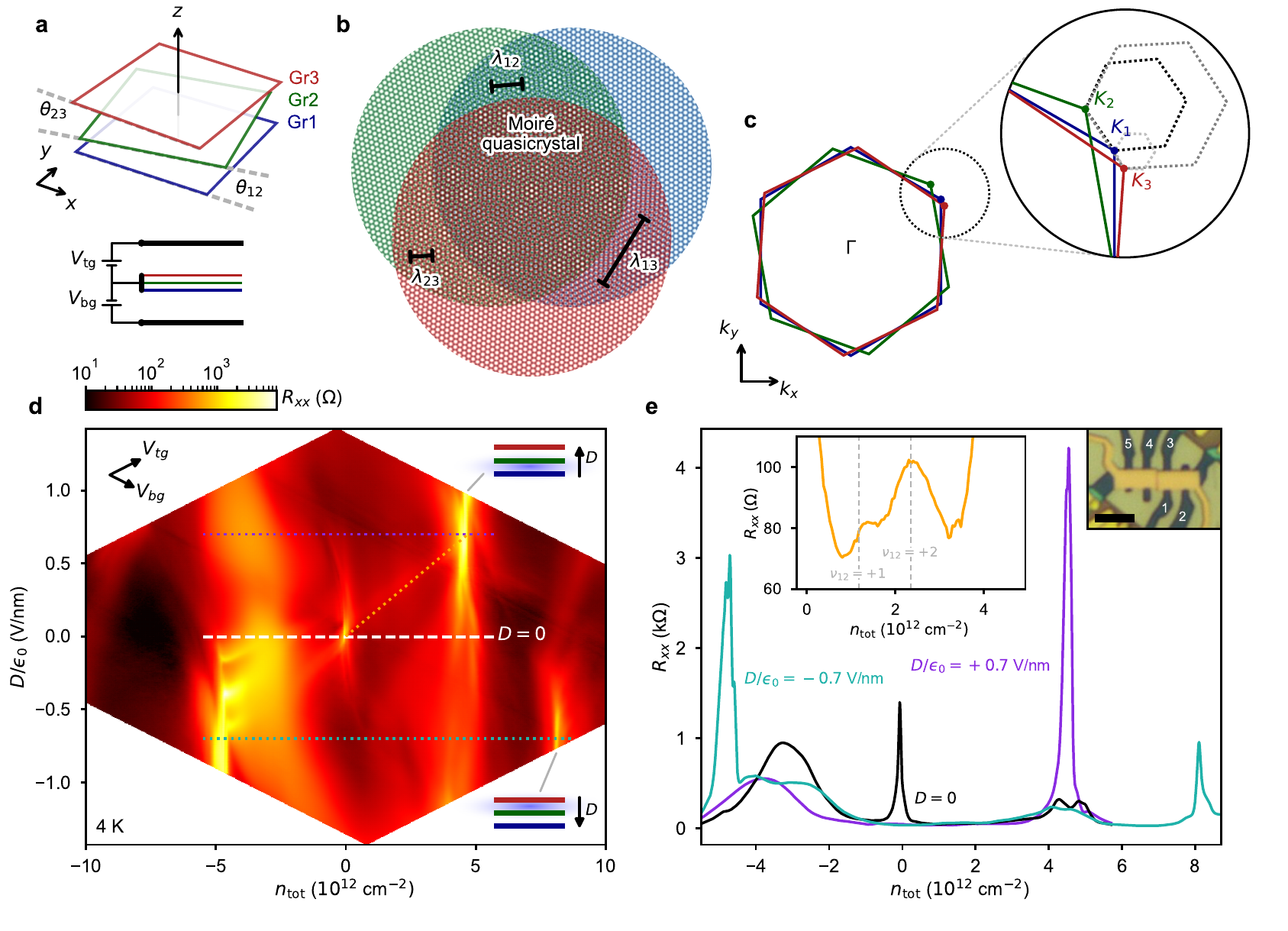}
    \caption{\figtitle{Realization of a \moire quasicrystal in two-angle TTG.}
    \panel{a} Schematic of the twisted trilayer structure, comprising three layers of graphene (Gr1, Gr2, Gr3) rotated by alternating unequal relative twist angles, $\theta_{12}$ and $\theta_{23}$.
    Bottom: circuit diagram of twisted trilayer graphene surrounded by two hexagonal boron nitride dielectric layers (not drawn) and top and bottom gate electrodes kept at potentials $V_\text{tg}$, $V_\text{bg}$ relative to TTG.
    \panel{b} Each pair of layers $i,j$ gives rise to \moire length $\lambda_{ij}$. 
    Their overlap gives rise to a quasiperiodic \moire pattern.
    \panel{c} The rotated Brillouin zones of the three monolayers result in three misaligned Dirac points ($K_i$) and two comparable pairwise mini-Brillouin zones.
    \panel{d} Longitudinal resistance $R_{xx}$ versus total carrier density $n_\text{tot}$ and electric displacement field $D$, measured at temperature $T = \SI{4}{K}$ and zero magnetic field.
    Insets illustrate situations where one of the layers is charge neutral and the remaining pair $i,j$ is at full-filling, $n_{\text{s},ij}$.
    \panel{e} Traces of $R_{xx}$ versus $n_\text{tot}$ at fixed $D$ showing asymmetry upon reflection through $D=0$, evidence of broken mirror symmetry due to $\theta_{12} \neq -\theta_{23}$.
    Middle inset shows two $R_{xx}$ peaks near $\nu_{12}=1,2$ in a trace along $n_3 \approx 0$ (dashed orange in \panel{d}), evidence of broken-symmetry states.
    All $R_{xx}$ data were measured between leads 1 and 2 unless stated otherwise (right inset, scale bar \SI{3.5}{\micro\meter}).
    }
    \label{fig:setup}
\end{figure*}

\section*{Realization of a \moire quasicrystal}
To explore the electronic properties of a system in this broad class of structures, we constructed a two-angle TTG system with two alternating and unequal twist angles, $0<\theta_{12}\neq-\theta_{23}>0$, and measured its four-terminal resistance (Fig.~\ref{fig:setup}\panel{a}, Methods~\ref{ssec:fab}).
The three incommensurate \moire lattices, defined by the pairs of layers $i$ and $j$, form a MQC (Fig.~\ref{fig:setup}\panel{b,c}, Fig.~\ref{fig:sos}\panel{f}).
We define the density of four electrons for each \moire unit cell as `full-filling', $n_{\text{s},ij}=4/A_{ij}$, where $A_{ij}\approx\sqrt{3} a^2/(2\theta_{ij}^2)$ is each unit cell area and 4 accounts for spin and valley degeneracy of the parent graphene layers.

In systems with mirror reflection symmetry, $z \rightarrow -z$, such as magic-angle TTG \cite{park2021tunable,hao2021electricfield}, transport properties are invariant under inversion of the electric displacement field, $D \rightarrow -D$.
Our system, in contrast, displays strong asymmetry with respect to $D \rightarrow -D$ (Fig.~\ref{fig:setup}\panel{d}-\panel{e}) due to its unequal twist angles.

\begin{figure*}
    \centering
    \includegraphics[width=6.75in]{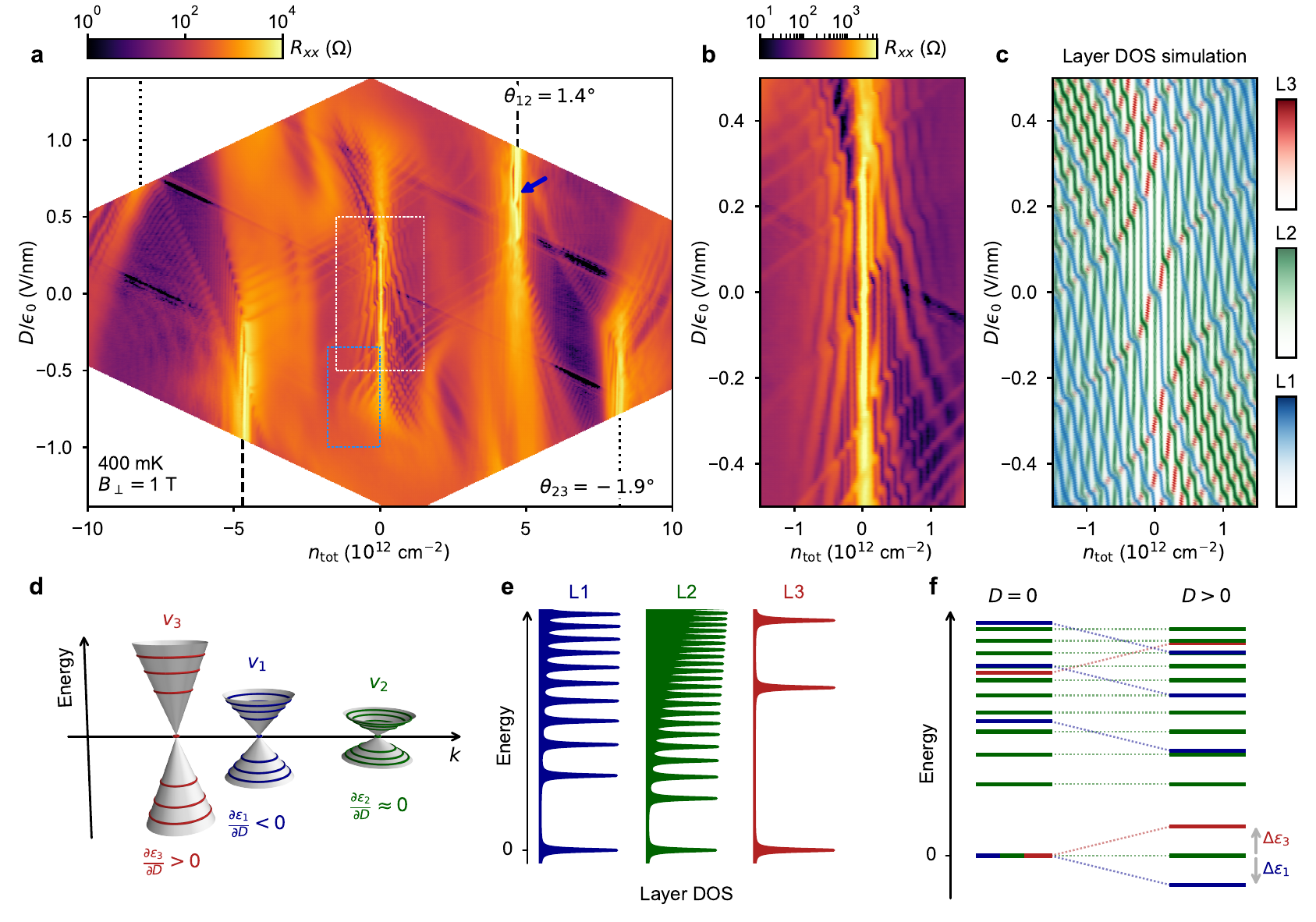}
    \caption{\figtitle{Evidence of periodic and quasiperiodic electronic regimes.}
    \panel{a} $R_{xx}$ measured at fixed perpendicular magnetic field, $B_\perp=\SI{1}{T}$, and temperature $T=\SI{400}{mK}$, versus $n_\text{tot}$ and $D$.
    Black dashed and dotted lines indicate $R_{xx}$ peaks associated with full-filling of the two pairwise \moire lattices, indicating twist angles $\theta_{12}=\SI{1.42\pm 0.07}{\degree}$ and $\theta_{23}=\SI{-1.88\pm 0.08}{\degree}$, respectively.
    Blue arrow indicates the alignment of the charge neutral point of layer~3 with full-filling of \moire of layers 1 and 2, $n_\text{s,12}$, generating a resistance peak.
    \panel{b} Detail of the dashed white box region in \panel{a} showing three sets of Landau levels with different slopes.
    \panel{c} Simulation of the DOS for three Dirac cones with layer hybridization giving rise to renormalized Fermi velocities $(v_1,v_2,v_3) = (0.51,0.2,1)v_3$ at $B_\perp=\SI{1}{T}$.
    Displacement field is accounted for by adding to each Landau level spectrum a linear energy shift $\Delta\varepsilon_{i} = \alpha_i D$ for cones $i=\lbrace 1,2,3\rbrace$.
    The three color ranges indicate the partial DOS on the three effective layers, $\text{L1,L2,L3}$.
    \panel{d} Schematic Dirac cones with Fermi velocities $v_i$ and rates of potential shift, $\alpha_i = \partial \varepsilon_i / \partial D$.
    \panel{e} Example DOS from each effective layer for $D=0$, with energy spacings differing between layers due to different Fermi velocities $v_i$.
    \panel{f} Applying nonzero $D$ shuffles the sequence of Landau levels due to the different potential shifts, $\Delta \varepsilon_i$.
    }
    \label{fig:LLs}
\end{figure*}

In the presence of a perpendicular magnetic field, $B_\perp=\SI{1}{T}$, we observe different sets of Landau level (LL) features that depend strongly on $D$ (Fig.~\ref{fig:LLs}\panel{a},\panel{b}).
This is in contrast to highly coupled multilayer systems like magic-angle TBG and TTG (magic-angle graphene), Bernal bilayer graphene, or rhombohedral trilayer graphene, where the LLs exhibit little $D$-field dependence \cite{hunt2017direct,yankowitz2019tuning,zhou2021halfandquarter}.
It suggests that the electronic states in our system possess layer character.
Accordingly, we define effective-layer-resolved carrier densities, $n_i$, for layers $i=\lbrace1,2,3\rbrace$, where the total (electron) carrier density is $n_\text{tot} = \sum_i n_i$.

In the absence of layer~3, upon doping the system such that $n_\text{tot} = n_1 + n_2 = n_{\text{s},12}$, the \moire unit cell defined by layers 1 and 2 holds four electrons and the Fermi energy enters a gap, resulting in an insulating state with large $R_{xx}$. 
However, the presence of layer~3 adds a parallel conducting channel, significantly diminishing the resistive peak.
By applying positive $D$ we deplete layer~3 so that $n_3 \approx 0$ and tune $n_\text{tot}\approx n_1+n_2=n_{\text{s},12}$ (\extlbl~\ref{fig:CNP_traces}\panel{a}) to recover the resistive state.
We find a sharp $R_{xx}$ peak at $n_\text{tot} = n_{\text{s},12}=\SI{4.7e12}{cm^{-2}}$ (blue arrow in Fig.~\ref{fig:LLs}\panel{a}) from which we extract $|\theta_{12}|=\SI{1.42}{\degree}$ (Methods~\ref{ssec:theta12}).

\begin{figure*}
    \centering
    \includegraphics[width=6.75in]{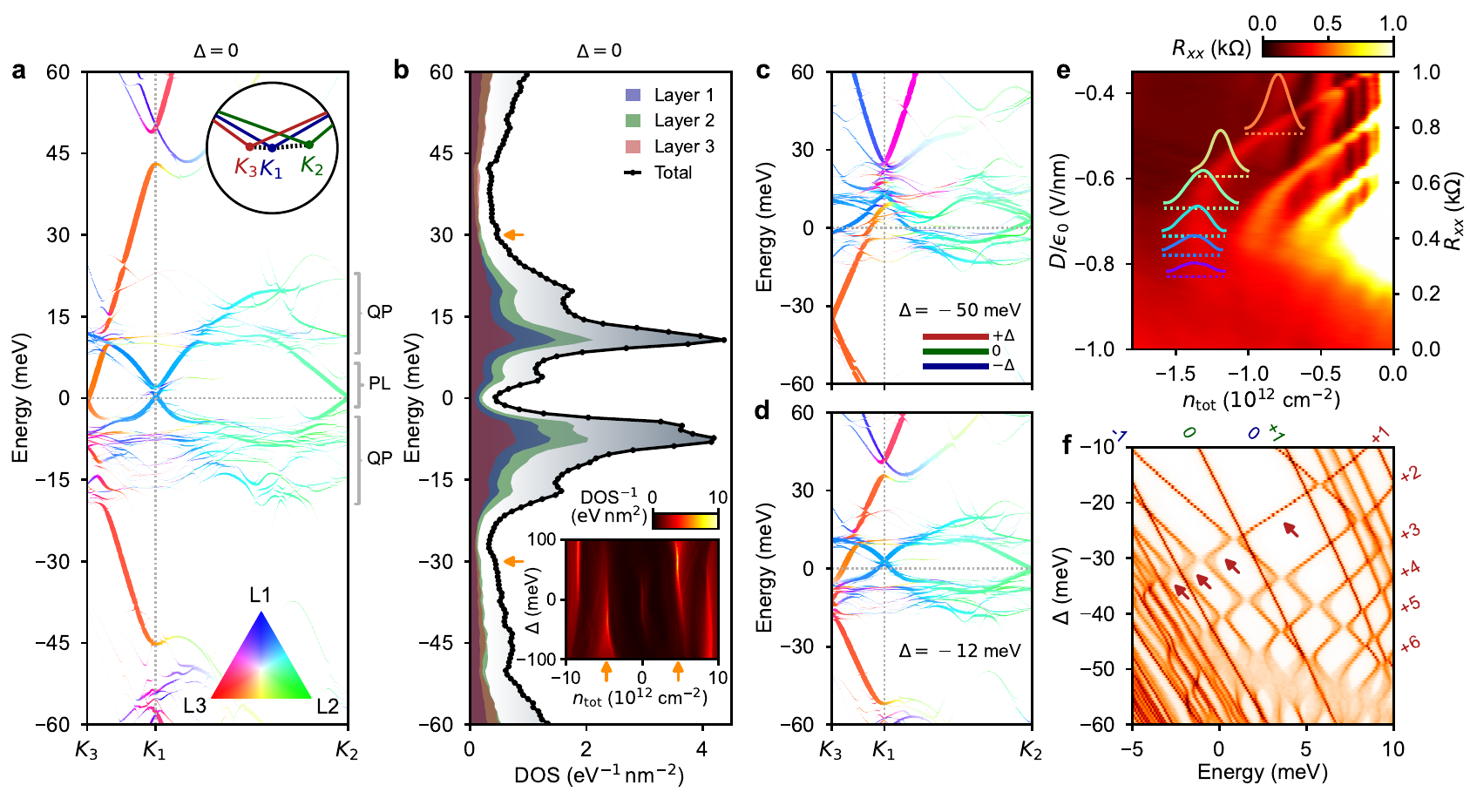}
    \caption{\figtitle{Consequences of \moire quasiperiodicity.}
    \panel{a} Spectral function calculated along a path between the layer $K$-points (see upper inset).
    Point size indicates the total spectral weight of each eigenstate at $\mathbf{k}$, while color represents the relative weight on the three layers according to the color triangle (bottom inset; Methods~\ref{ssec:SF}).
    PL -- periodic-like regime; QP -- quasiperiodic regime.
    \panel{b} Layer-resolved DOS and total DOS computed by integrating the spectral function over all $k_x$,$k_y$.
    Inset shows inverse total DOS versus $n_\text{tot}$ and $\Delta$.
    Orange arrows indicate $n_\text{tot}=n_\text{s,12}$.
    \panel{c} Same as \panel{a} for layer potential $\Delta = \SI{-50}{meV}$ (see inset), corresponding to $D/\epsilon_0 \approx \SI{-1}{V/nm}$.
    $\varepsilon=0$ is fixed to charge neutrality.
    \panel{d} Same as \panel{c} for $\Delta = \SI{-12}{meV}$, showing relative shifts of Dirac nodes near the $K$-points.
    \panel{e} Measured $R_{xx}$ versus $n_\text{tot}$ and $D$ from the dashed blue box in Fig.~\ref{fig:LLs}\panel{a}. 
    Line traces of $R_{xx}$ (right axis), following the $N=+1$ LL of the fast Dirac cone, are superimposed for selected $D/\varepsilon_0$ values (indicated by horizontal dashed lines), and shifted for clarity.
    Linear fits were subtracted from the cuts.
    \panel{f} DOS at $B_\perp=\SI{1}{T}$ calculated for a periodic approximant (Methods~\ref{ssec:SF_finite_B}) that captures the essence of the quasiperiodic structure. 
    Arrows highlight the energy broadening of LL $N=+1$ of the fast Dirac cone upon transitioning from the periodic-like regime to the quasiperiodic regime.
    LL indices are indicated (colors represent dominant layer character).
    }
    \label{fig:bands}
\end{figure*}

We observe another resistive peak at a higher $n_\text{tot}$ and opposite $D$ field (Fig.~\ref{fig:LLs}\panel{a}), associated with $n_{\text{s},23}$, indicating $|\theta_{23}|=\SI{1.88}{\degree}$ (see also Fig.~\ref{fig:bands}\panel{b}, inset; Methods~\ref{ssec:theta23}).
The two resistive peaks at different $n_\text{tot}$ indicate the presence of multiple \moire unit cells, a direct signature of a MQC.
We assembled the layers with angles of opposite signs, therefore $(\theta_{12},\theta_{23})=(\SI{1.42\pm 0.07}{\degree},-\SI{1.88\pm 0.08}{\degree})$ (Methods~\ref{ssec:twist}).

\section*{low energy phenomenological model}
Since the three graphene layers are misoriented, we expect the low energy dispersion to contain three Dirac cones protected by an approximate $C_{2z}T$ symmetry, with Fermi velocities renormalized by the interlayer tunneling \cite{amorim2018electronic,zhu2020twistedtrilayer}.
Indeed, Figs.~\ref{fig:LLs}\panel{a,b} show three sets of LLs that differ by their relative slopes in the $n_\text{tot}$-$D$ plane, with non-uniform density separation between LLs within each set.
We understand this structure using a simple low-energy phenomenological model consisting of three Dirac cones with renormalized Fermi velocities, $v_i$, for $i=\lbrace1,2,3\rbrace$ (Fig.~\ref{fig:LLs}\panel{d}).
At $D=0$ the LL energies of the three Dirac cones are $\varepsilon_{N,i}=v_i \operatorname{sgn}(N)\sqrt{2\hbar eB|N|}$, where $N$ is the LL index (Fig.~\ref{fig:LLs}\panel{e}) \cite{castroneto2009electronic}.
The lowest velocity cone develops an energy-dense sequence of LLs, while the faster cones generate sparser spectra (Figs.~\ref{fig:LLs}\panel{d,e}).
Upon increasing $n_\text{tot}$, the Fermi energy passes through LLs belonging to different cones in a sequence determined by the energy ordering of the LLs (Fig.~\ref{fig:LLs}\panel{f}, left).
For $D\neq 0$, we include linear shifts in the energies of the Dirac cones, $\Delta \varepsilon_{N,i} = \alpha_i D$, reflecting the $D$-field induced potential imbalance across the three Dirac cones due to their partial layer polarization.
As a result, the LL sequence is shuffled (Fig.~\ref{fig:LLs}\panel{f}, right).
Figure~\ref{fig:LLs}\panel{c} shows the calculated density of states (DOS) of the LLs according to the above phenomenological model, assuming a degeneracy of $4B/\phi_0$ for all LLs (Methods~\ref{ssec:phenomenological}), where $\phi_0=h/e$ is the flux quantum, and $h$ and $e$ are Planck's constant and the elementary charge, respectively.
We tune the parameters $v_i$ and $\alpha_i$ to match the measured resistance map at low densities and allow for small quadratic terms in the Dirac dispersions (Methods~\ref{ssec:phenomenological}).
The velocities $v_i$ can be extracted up to an overall factor using this method, yielding the velocity ratios $v_1/v_3 = 0.51$ and $v_2/v_3 = 0.20$.
These ratios corroborate the extracted twist angles (Methods \ref{ssec:perturbative}) and imply that all three monolayers are partially hybridized, while also retaining layer character.
Calculations of the electronic structure confirm this, as we describe below.

\section*{Electronic structure}
While one cannot construct Bloch bands without periodicity, the spectral function (SF), or the probability of an energy eigenstate to appear at $(\varepsilon,\mathbf{k})$, remains well-defined.
We compute the SF for our system using a momentum-space method valid at arbitrary twist angles \cite{koshino2015interlayer,massatt2017incommensurate,amorim2018electronic,zhu2020twistedtrilayer}.
We model the individual monolayers using a tight binding dispersion and introduce interlayer coupling through generalized umklapp scattering in momentum space using the experimentally extracted twist angles.
We fine-tune the model parameters using our experimentally extracted Fermi velocity ratios (Methods~\ref{ssec:SF}).

Figure~\ref{fig:bands}\panel{a} shows the calculated SF along the path $K_3 \rightarrow K_1 \rightarrow K_2$ (Fig. ~\ref{fig:bands}\panel{a}, top inset).
Each curve comprises points at different $(k,\varepsilon)$.
The size of each point represents the total projection onto states from all three layers at that momentum, and the color represents the relative weight on each layer (Fig.~\ref{fig:bands}\panel{a}, bottom inset).
At low energies we find three Dirac cones with different Fermi velocities centered on the $K$ points of the three monolayers.
In this periodic-like regime (PL in Fig.~\ref{fig:bands}\panel{a}), the states appear continuous and plane-wave like (spectral weight concentrated on a single $k$ point, indicated by large point size), similar to periodic bands.
As expected, the fast, medium, and slow velocity cones (orange, $K_3$; light blue, $K_1$; light green, $K_2$, respectively) have spectral weights mostly on the top, bottom, and middle graphene layers, respectively, however, they also show significant hybridization.

The full SF shows a group of weakly dispersing states (light green and blue, $|\varepsilon| \lesssim \SI{20}{meV}$) reminiscent of the flat bands in magic-angle graphene systems \cite{khalaf2019magic}, surrounded by soft gaps induced by \moire coupling of layers 1 and 2 (reduced DOS in Fig.~\ref{fig:bands}\panel{b}).
In contrast to magic-angle graphene, at energies above the well-defined Dirac nodes, a quasiperiodic regime emerges (QP in Fig.~\ref{fig:bands}\panel{a}).
There, the quasicrystalline order forms a dense set of avoided crossings.
Additionally, the eigenstates are not plane waves, evident by the small point sizes in the SF.
We note that the quasiperiodic and periodic-like regimes can appear at different energies for different MQCs (\supplbl~\ref{fig:other-angles}).

Figures~\ref{fig:bands}\panel{c,d} show the effect of interlayer potential asymmetry on the SF (see Supplementary Video 1 for the full sequence).
We simulate the effect of $D$-field by shifting the electric potentials of the outer layers by $\pm \Delta$ (Fig.~\ref{fig:bands}\panel{c}, inset), where $\Delta = \SI{50}{meV}$ corresponds to $D/\epsilon_0 \approx \SI{1}{V/nm}$ (Methods~\ref{ssec:electrostatics}).
Figure~\ref{fig:bands}\panel{d} shows that $\Delta$ shifts the energy of each Dirac node, consistent with the phenomenological model employed in Figs.~\ref{fig:LLs}\panel{c}-\panel{f}.

The calculated DOS (Fig.~\ref{fig:bands}\panel{b}) shows two peaks, reminiscent of the Van Hove singularities of the flat bands in magic-angle graphene, that coincide with the quasiperiodic regime.
The calculated inverse DOS, plotted versus $n_\text{tot}$ and $D$ (Fig.~\ref{fig:bands}\panel{b}, inset), reveals peaks at $\pm n_{\text{s},12}$ and $\pm n_{\text{s},23}$, consistent with the resistance peaks used to extract the twist angles (Methods~\ref{ssec:DOS}).

\section*{Transport signature of the periodic-quasiperiodic transition}
Our MQC provides an opportunity to explore magnetic oscillations in a tunable quasiperiodic system \cite{spurrier2018semiclassical}.
In the periodic-like regimes, we observe sharp LLs, characteristic of periodic lattices under small magnetic fields.
In contrast, in the quasiperiodic regimes, characterized by dense avoided crossings in the SF, we observe reduced LL visibility in the $R_{xx}$ data.
This is accounted for by LL broadening, reflected in our DOS calculation under magnetic field (Methods~\ref{ssec:SF_finite_B}).
We observe reduced LL visibility in several regions in $n$-$D$ space:
(i) LLs from the two slow Dirac cones fade away upon entering the quasiperiodic regime ($|n_\text{tot}| \gtrsim \SI{2e12}{cm^{-2}}$ in Fig.~\ref{fig:LLs}\panel{a}; QP in Fig.~\ref{fig:bands}\panel{a}; \extlbl~\ref{fig:SF-finite-B}\panel{b}).
(ii) A strong quasiperiodic regime exists at high $D$ around charge neutrality resulting in complete absence of $R_{xx}$ oscillations there ($|D/\epsilon_0| \gtrsim \SI{0.9}{V/nm}$ in Fig.~\ref{fig:LLs}\panel{a}, $|\Delta| \gtrsim \SI{50}{meV}$ in Fig.~\ref{fig:bands}\panel{c}, and \extlbl~\ref{fig:SF-finite-B}\panel{c}).
(iii) Perhaps the most striking quasicrystalline feature is the continuously diminishing $R_{xx}$ peak height observed when tracing, for example, the $N=+1$ LL of layer~3 as $|D|$ is increased (Fig.~\ref{fig:bands}\panel{e}).
As the fast band energy is shifted by $D$, the LL aligns with regions of increasingly high DOS in the flat band.
The phase space for umklapp scattering grows, more heavily reconstructing the fast cone (Fig.~\ref{fig:bands}\panel{d}) and broadening the LL (arrows in Fig.~\ref{fig:bands}\panel{f}).
This unusual change in the width of the LLs at fixed magnetic field is not expected from any disorder mechanism.
Rather, it is a consequence of continuous modulation of the electronic structure arising from quasiperiodicity (Methods~\ref{ssec:SF_finite_B}).

\begin{figure*}
    \centering
    \includegraphics[width=6.75in]{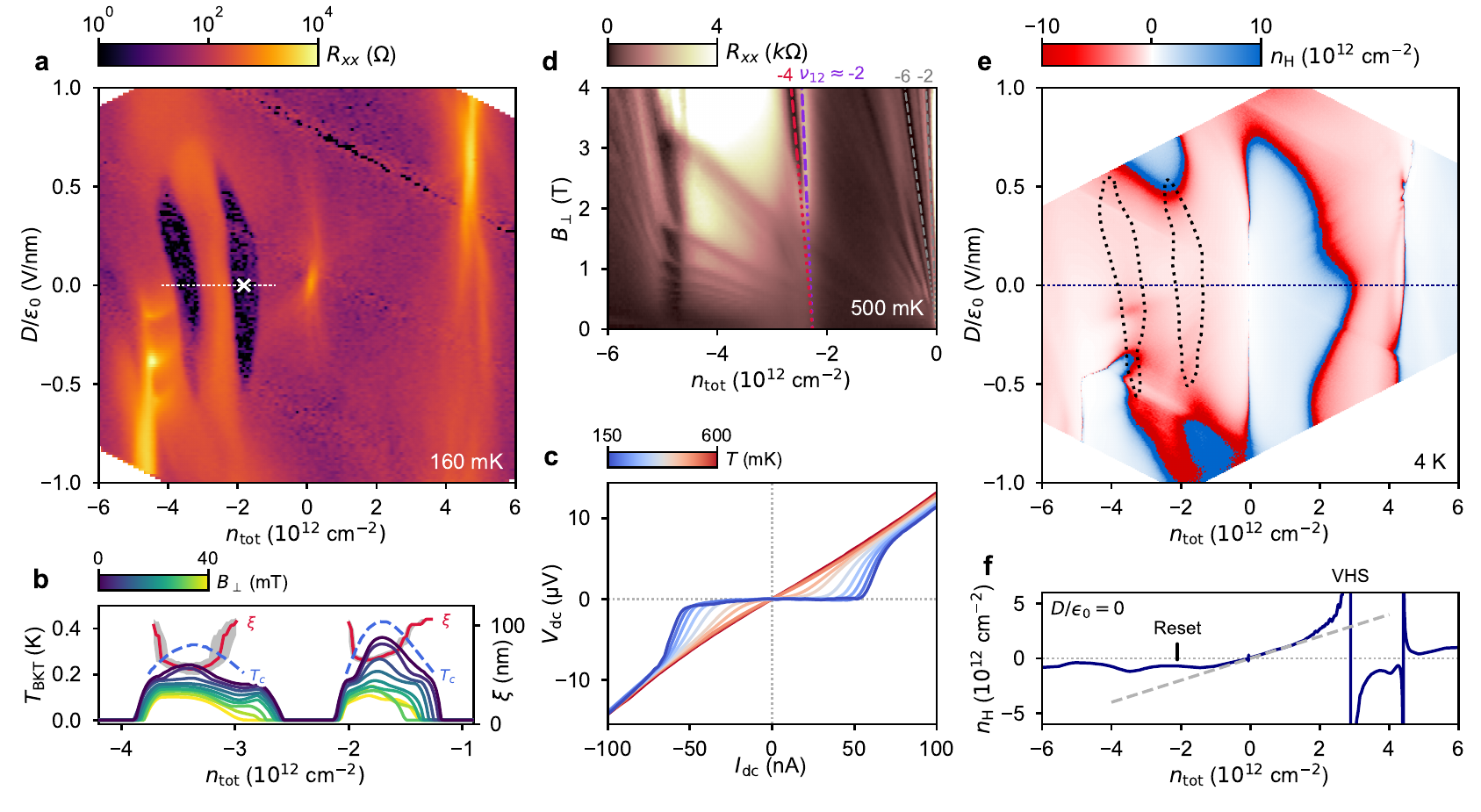}
    \caption{\figtitle{Strong electronic interactions and superconductivity.}
    \panel{a} $R_{xx}$ versus $n_\text{tot}$ and $D$ measured at $T = \SI{160}{mK}$ showing two superconducting pockets, separated by a resistive region that coincides with a flavor-symmetry-breaking phase transition.
    \panel{b} Berezinskii–Kosterlitz–Thouless superconducting transition temperature, $T_\text{BKT}$, and critical temperature, $T_c$, at 50\% of the normal resistance (left axis), and Ginzburg-Landau coherence length, $\xi$ (right axis), along the white dotted line in \panel{a}.
    We extract $\xi$ from the $B_\perp$-field dependence of $T_\text{c}$.
    Gray shading indicates error bars corresponding to $T_\text{c}$ between 40\% and 60\% of the normal resistance.
    \panel{c} Current-voltage characteristics at various temperatures, measured at the position of the white cross in \panel{a}.
    \panel{d} $R_{xx}$ versus $n_\text{tot}$ and $B$, measured between contacts 4 and 5 (Fig.~\ref{fig:setup}\panel{e}, inset), showing LLs emanating from $\nu_{12} \approx -2$, indication of a flavor-symmetry-breaking phase transition.
    \panel{e} Hall density, $n_\text{H}$, measured at $T=\SI{4}{K}$ and $B_\perp=\pm\SI{0.2}{T}$, showing a Van Hove singularity on the electron side, near $n_\text{tot} = \SI{2e12}{cm^{-2}}$.
    The absence of a Van Hove singularity on the hole side, $n_\text{tot}<0$, is a consequence of the phase transition.
    \panel{f} $n_\text{H}$ versus $n_\text{tot}$ at $D=0$ (along dashed line in \panel{e}), showing the Van Hove singularity at $n_\text{tot}\approx \SI{2.8e12}{cm^{-2}}$, and the phase transition near $\nu_{12} = -2$.
    }
    \label{fig:SC}
\end{figure*}

\section*{Strong electronic interactions and superconductivity}
In addition to quasiperiodicity, the system exhibits phenomena beyond single-particle physics.
We observe a spontaneous flavor-symmetry-breaking phase transition \cite{wong2020cascade,zondiner2020cascade} at moderate $D$ fields indicated by a set of LLs originating from $\nu_{12} \approx -2$, or two holes per \moire unit cell formed by layers 1 and 2 (Fig.~\ref{fig:SC}\panel{d} and Methods~\ref{ssec:DR}), accompanied by a drop (reset) in the Hall density, $n_\text{H}$ (Figs.~\ref{fig:SC}\panel{e,f}).
On the electron-doped side, increased resistance around integer fillings, $\nu_{12}=+1,+2$ (and approximate neutrality of layer~3), suggests the formation of correlated states \cite{cao2018correlated} at these densities due to electronic interactions (Fig.~\ref{fig:setup}\panel{e}, inset).

At low temperatures on either side of the flavor-symmetry-breaking phase transition, we observe two superconducting 
pockets of zero resistance (Fig.~\ref{fig:SC}\panel{a}), with maximal Berezinskii–Kosterlitz–Thouless transition temperature $T_{\text{BKT}}\approx \SI{300}{mK}$ (Fig.~\ref{fig:SC}\panel{b}) and non-linear current-voltage characteristics (Fig.~\ref{fig:SC}\panel{c}).
The two superconducting pockets are separated in density by a metallic region near the flavor-symmetry-breaking phase transition.
In contrast to magic-angle graphene, where broken flavor symmetry appears to be required for superconductivity, the right superconducting pocket in our system is hosted by a flavor-symmetric state.
Moreover, the left superconducting pocket may be related to the right one, as they appear at approximately the same filling fraction per flavor (Methods~\ref{ssec:SC-analysis}).

Importantly, both superconductivity and the flavor-symmetry-broken phase appear at densities where the quasicrystalline nature of the system is especially pronounced, as indicated by the absence of LLs from Dirac cones 1 and 2 (see outline of superconducting pockets in \extlbl~\ref{fig:CNP_traces}\panel{a}) as well as by the SF calculations that show dense avoided crossings at similar fillings (QP in Fig.~\ref{fig:bands}\panel{a}).

Emergent superconductivity has been previously reported in \moire periodic systems such as magic-angle TBG and TTG \cite{cao2018unconventional, park2021tunable, hao2021electricfield}, however, the twist angles in our system are significantly outside the regimes of existing magic-angle superconductivity.
Furthermore, magic-angle TBG and TTG are both single-angle systems (akin to \figlbl~\ref{fig:sos}\panel{a},\panel{c}) with well-defined \moire periodic bands hosting the superconductivity.
In the latter case, $\theta_{12} = -\theta_{23}=\theta$, and $\theta_{13}=0$, which allows mapping the band structure in magic-angle TTG to magic-angle TBG flat bands with a superimposed Dirac cone \cite{khalaf2019magic}.
This mapping is not possible in our system due to the unequal twist angles, which yields a qualitatively different electronic structure.
Thus, the electronic structure and symmetries in our MQC are fundamentally different from magic-angle systems, and the superconducting state may thus be distinct.

While the exact nature and origin of the superconductivity in our system is not known, our estimate of the ratio $T_\text{c}/T_\text{F} \approx 0.008$ indicates the superconductivity approaches the strong-coupling regime \cite{cao2018unconventional}. 
Here, $T_c \approx \SI{0.4}{K}$ is the critical temperature extracted at 50\% of the normal resistance (Fig.~\ref{fig:SC}\panel{b}), $T_\text{F} = n_\text{tot}/k_\text{B}\rho_\text{F} \approx \SI{50}{K}$ is the estimated Fermi temperature, and $\rho_\text{F}$ is the DOS at the Fermi energy.
In the absence of relevant magnetic oscillations in the superconducting state (\extlbl~\ref{fig:CNP_traces}\panel{a}), we use the calculated DOS value, $\rho_\text{F} = \SI{4}{eV^{-1}nm^{-2}}$ (Fig.~\ref{fig:bands}\panel{b}).
This high $T_\text{c}/T_\text{F}$ ratio is corroborated by the relatively low ratio $\xi/d \approx 9$.
Here, $\xi\approx \SI{70}{nm}$ is the Ginzburg-Landau coherence length (Figure \ref{fig:SC}\panel{b}), and $d = n_\text{tot}^{-1/2} \approx \SI{7.5}{nm}$ is the interparticle distance.
This places our system between the weak-coupling Bardeen-Cooper-Schrieffer (BCS) regime (typically $100 < \xi/d < 10^4$), and the ultra-strong coupling regime ($\xi/d \approx 1$, at the crossover from BCS to Bose-Einstein condensation), close to magic-angle graphene \cite{cao2018unconventional,park2021tunable}.

In common with all known robust superconducting \moire graphene systems \cite{cao2018unconventional,park2021tunable,hao2021electricfield,park2022robust,zhang2022promotion,burg2022emergence}, our system possesses an approximate $C_{2z}T$ symmetry, a proposed requirement for strong-coupling superconductivity in \moire systems \cite{khalaf2021charged}.
However, quasiperiodicity in our system may further constrain the allowed order parameter symmetries.
For instance, nodal intravalley pairing may be suppressed by the quasiperiodic scattering, similar to the effect of disorder in unconventional superconductors \cite{mackenzie1998extremely} but with different symmetries and on larger length scales that correspond to intravalley processes.
This could be in striking contrast to spectroscopic evidence of nodal pairing in magic-angle graphene \cite{kim2022evidence,oh2021evidence}, though further theoretical and experimental investigations are required to explore the connection between superconducting phases in \moire periodic and MQC graphene structures.

\section*{Conclusions}
Two-angle twisted trilayer graphene combines the flat band physics and tunability of \moire systems with the unique nature of quasiperiodic long-range order, suggesting new directions for \moire and quasicrystal investigations alike.
Relative to the limited tunability and engineering challenges of conventional metallic-alloy quasicrystals, MQCs can be easily assembled from simple building blocks with many tunable parameters.
These include carrier density, electric displacement and magnetic fields, and importantly, the \moire quasicrystalline structure itself by controlling the twist angles.
The use of other materials will greatly expand the class of \moire quasicrystals, leading to new quasiperiodic systems displaying a variety of electronic properties and phenomena beyond what is reported here.
We anticipate that this new class of \moire quasicrystals will provide an experimental platform for exploring open questions in quasiperiodic systems, both at the single-particle level and in the strongly-interacting regime.

\clearpage

\ExtendedData
\section*{Methods}

\subsection{Device fabrication}\label{ssec:fab}
The device consists of an hBN-encapsulated twisted trilayer graphene stack with metallic top and bottom gates, fabricated using a combination of cut-and-stack and hot release methods.
Graphene and hexagonal boron nitride (hBN) were exfoliated onto \ce{SiO2}/\ce{Si} substrates, and desired flakes were selected using an optical microscope.
The heterostructure was assembled using a polymer-based dry transfer technique.
A glass slide with a poly(bisphenol~A carbonate) (PC) film covering a polydimethylsiloxane (PDMS) block was mounted onto the micro-positioning stage of a homebuilt transfer setup, and used to sequentially pick up the van der Waals flakes.
First, a bottom hBN on a metal back gate was prepared.
The back gate was formed by thermal deposition of \SI{2}{nm} \ce{Cr} / \SI{27}{nm} \ce{PdAu} (60\% Au, 40\% Pd) onto a Si substrate, then annealed in \ce{H2} and Ar at \SI{300}{\degree C}.
Next, a suitable bottom hBN flake was deposited onto the back gate and the PC film was dissolved in a chloroform bath.
The chip was annealed in forming gas at \SI{350}{\degree C}, followed by an atomic force microscopy (AFM) tip cleaning in contact mode to ensure any polymer residues were removed.
For the top stack, first the top hBN was picked up by heating the substrate to \SI{90}{\degree C}.
The hBN was then used to pick up, at room temperature, the first of the three pieces cut from a single monolayer graphene flake.
Next, the chip with the two remaining graphene flakes was rotated to an angle close to \SI{+1.6}{\degree}, and the second graphene piece was picked up.
The third graphene piece was rotated back by an angle close to \SI{-1.6}{\degree}, and then picked up.
As is common in twisted \moire stacks, the angles relaxed to their final positions during subsequent fabrication steps.
This four-layer stack was deposited onto the previously prepared bottom hBN on metallic back gate by melting the PC at \SI{170}{\degree C}, and the PC film was dissolved in a chloroform bath. 

The final stack was inspected using dark field optical microscopy and AFM to select bubble-free regions in which to define Hall bars.
All fabrication steps involved patterning the heterostructure using a polymethyl-methacrylate (PMMA) resist mask and electron beam lithography (EBL).
The first step was an etch through the entire stack using reactive ion etching (RIE) in an Ar, \ce{CHF3}, \ce{O2} plasma environment.
This etch was performed away from the region of interest, intended to minimize the twist-angle relaxation in subsequent steps.
Next, we patterned 1D edge contacts onto the graphene regions that extended over the back gate.
The heterostructure regions exposed by the lithographic mask were etched, followed by a deposition of the contact electrodes, consisting of a \SI{3}{nm} \ce{Cr} sticking layer and \SI{90}{nm} of \ce{Au}, performed using a tilted rotating stage in a thermal evaporator.
Liftoff was performed in acetone at room temperature.
Next, the top gate was patterned with another series of EBL, thermal evaporation and liftoff.
Then, the final Hall bar geometry was defined using EBL followed by an RIE etch.

\begin{figure}
    \centering
    \includegraphics[width=88mm]{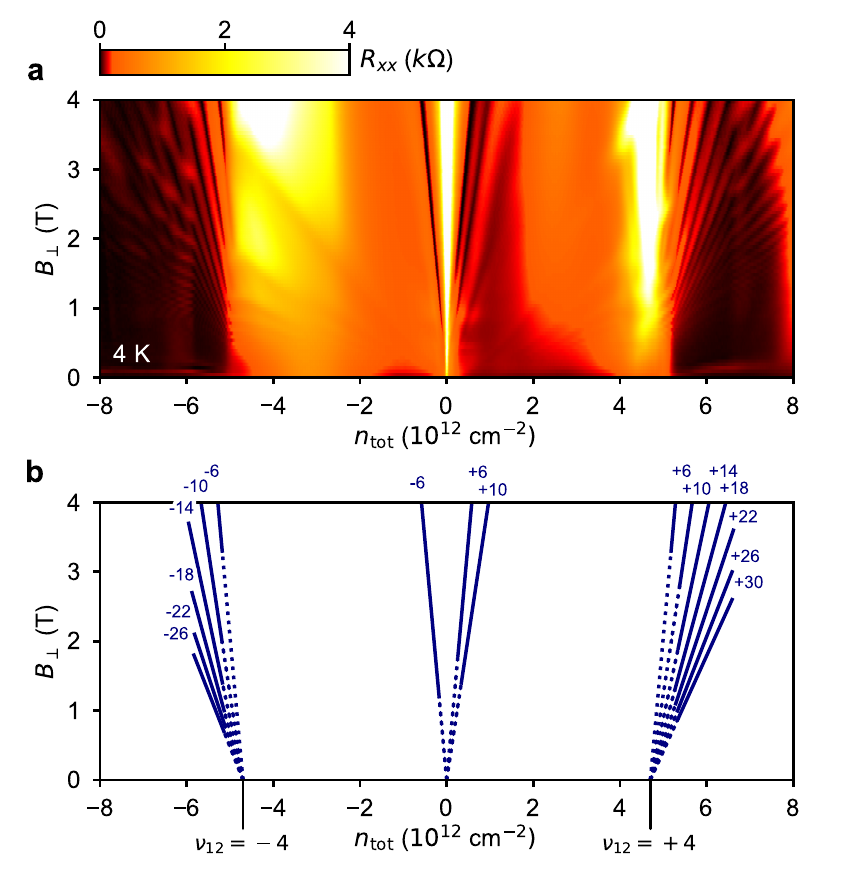}
    \caption{\figtitle{Extraction of twist angle, $\theta_{12}$.}
    \panel{a} Measured $R_{xx}$ versus $n_\text{tot}$ and perpendicular field $B_\perp$ at temperature $T=\SI{4}{K}$, with fixed $D=0$.
    Landau levels beyond full filling of the \moire unit cell formed by layers 1 and 2 ($n_{\text{s},12} = \SI{4.7e12}{cm^{-2}}$) emerge from $n_{\text{s},12}$.
    \panel{b} Extracted Landau levels used to calibrate the geometric capacitances (together with $n_\text{tot}$-$D$ behavior) and used to determine $n_{\text{s},12}$ ($\nu_{12}=\pm 4$).
    }
    \label{fig:fanfit}
\end{figure}

\subsection{Measurement setup}\label{ssec:measurement_setup}
We measured transport data in a dilution refrigerator with a base temperature of \SI{150}{mK}. 
We used top and bottom gate voltages ($V_\text{tg}$, $V_\text{bg}$) to independently control the total carrier density, $n_\text{tot} = (C_\text{bg}V_\text{bg} + C_\text{tg}V_\text{tg})/e$, and electric displacement field, $D = (C_\text{bg}V_\text{bg}-C_\text{tg}V_\text{tg})/2$, where $C_\text{tg}$ ($C_\text{bg}$) is the top-gate (back-gate) capacitance per unit area (Fig.~\ref{fig:setup}\panel{a}, bottom).
We measured the electrical resistance in a four-terminal Hall bar geometry (Fig.~\ref{fig:setup}\panel{e}, inset) using conventional lock-in techniques.
For the current-voltage characteristics (Fig.~\ref{fig:SC}\panel{c}) we used a home-built DC voltage source in series with a \SI{10}{\mega\ohm} resistor to current bias the sample, and measured the voltage using a digital multimeter, connected after a voltage pre-amplifier.

\subsection{Twist angle extraction}\label{ssec:twist}
Although certain twist angles are targeted during the stacking procedure, relaxation of the assembled layers causes the twist angles to change during subsequent fabrication steps.
The final twist angles must therefore be extracted from the measurements.
The standard method for extracting twist angles in \moire materials is to use the fact that band gaps (and thus insulating states) emerge at full filling of a \moire unit cell, at a superlattice density of $n_\text{s} = 4/A_\text{m}$, where the \moire unit cell area is given by $A_\text{m} \approx \sqrt{3}a^2 / (2\theta^2)$ for lattice constant $a$.
To obtain an accurate measure of $n_\text{s}$, we use the fact that LLs emerge from band edges at the superlattice density.
In this way, the convergence of multiple LLs as $B_\perp$ is reduced can be used to accurately determine $n_\text{s}$.
The density scale itself is calibrated by first estimating the geometric capacitances per unit area based on the parallel-plate formula, $C_i = \epsilon\epsilon_0 / d_i$, where $i$ labels the top or bottom gate hBN dielectric.
The hBN thicknesses $d_i$ are first estimated using atomic force microscopy.
The capacitances are then accurately calibrated using the fact that cyclotron gaps occur between Landau levels with integer slopes $t$ in the formula $n_t = t B/\phi_0$.

\subsubsection{Extraction of $\theta_{12}$}\label{ssec:theta12}
In our system the procedure is slightly more complex since there are two twist angles and many coexisting states in the spectral function, preventing global gap formation \cite{mora2019flatbands}.
Nevertheless, soft gaps (minima in the DOS) appear near $|\nu_{12}|=4$ from the $G_{12} = G_1 - G_2$ generalized umklapp scattering \cite{oka2021fractal}.
Thus, LLs from layer~3 that exist within the soft gap can still be used to extract the smaller of the two angles, $\theta_{12}$ (\extlbl~\ref{fig:fanfit}\panel{a}, $|n_\text{tot}| \gtrsim \SI{5e12}{cm^{-2}}$; Methods~\ref{ssec:SF}).
The reduced DOS from layers 1 and 2 above the edge of the high DOS energy band (flat quasi-band) is apparent from the vertical slopes acquired by LLs from layer~3 (see \extlbl~\ref{fig:CNP_traces}\panel{a}, $|\nu_{12}|>4$ and Section~\ref{ssec:layer_resolved}).
As the magnetic field is lowered to $B_\perp=0$ the density of the layer~3 LLs, $n=4B/\phi_0$, shrinks to zero.
Therefore the trajectory of the layer~3 fan at $n_\text{tot}>n_{\text{s},12}$ must converge to $n_\text{tot}=n_{\text{s},12}$ at $B_\perp=0$.
We find that these LLs emerge from $n_{\text{s},12} = \SI{4.7e12}{cm^{-2}}$ ($\nu_{12}=\pm 4$ in \extlbl~\ref{fig:fanfit}\panel{b}), yielding $\theta_{12} = \SI{1.42}{\degree}$ (see Section~\ref{ssec:theta_err} for error analysis).
We note that this method for estimating $\theta_{12}$ is possible because the LLs from layer 3 have high visibility above $n_{\text{s},12}$. The visibility is determined by the $\theta_{12}$ twist angle disorder \cite{uri2020mapping} which we estimate to be very low between the relevant contacts, $\delta \theta_{12} < 1 \%$. 

\subsubsection{Extraction of $\theta_{23}$}\label{ssec:theta23}
Dielectric breakdown sets limits on the gate voltages we can apply.
Therefore, the Landau level method cannot be easily applied to extract $\theta_{23}$.
Instead, we obtain $\theta_{23}$ from $R_{xx}$ features in the $(n_\text{tot},D)$ map at fixed field.
Once the geometric capacitances are calibrated using the procedure outlined in Section~\ref{ssec:theta_err}, we identify the density at which the peak in $R_{xx}$ occurs for $D<0$ in the vicinity of $n_\text{tot} = \SI{8e12}{cm^{-2}}$ at $B_\perp=\SI{1}{T}$ (Fig.~\ref{fig:LLs}\panel{a}).
We assign the center of the peak in resistance to be full filling for the \moire lattice defined by layers 2 and 3, $\nu_{23}=4$, and use this density, $n_{\text{s},23} = \SI{8.2e12}{cm^{-2}}$, to obtain $\theta_{23}$.
Within the limits of the gate voltages we can apply, we cannot reach neutrality of layer~1 while filling layers 2 and 3 to achieve $|\nu_{23}| = 4$.
This condition is expected to be fulfilled only at very high $D$ field due to the high Fermi velocity of layer~3 (see Fig.~\ref{fig:bands}\panel{c} and Supplementary Video 1).
However, our calculated inverse DOS shows that the peak in inverse DOS forms an approximately vertical feature in the $n_\text{tot}$-$\Delta$ plane (\extlbl~\ref{fig:idos}).
We use this fact to extract $n_{\text{s},23}$ from the peak in $R_{xx}$ accessible to us at moderate $D$ field.
In the inverse DOS map, calculated for $(\theta_{12},\theta_{23}) = (\SI{1.4}{\degree}, \SI{-1.9}{\degree})$, the peak appears at densities that imply $\theta_{23} = \SI{1.96}{\degree}\pm\SI{0.04}{\degree}$, with the lower value of $\SI{1.92}{\degree}$ appearing at moderate $\Delta$ values.
We therefore estimate that our method of extracting $\theta_{23}$ from the $R_{xx}$ peak systematically slightly overestimates $\theta_{23}$ by approximately $\SI{0.02}{\degree}$.
The capacitance error dominates over this error (see Section~\ref{ssec:theta_err}).

\subsubsection{Estimating twist angle errors}\label{ssec:theta_err}
The main source of error in estimating the twist angles is in the calibration of the capaticances $C_\text{tg}$, $C_\text{bg}$.
The ratio $C_\text{tg}/C_\text{bg}$ is calibrated so that the LLs in the $(n_\text{tot},D)$ map (Fig.~\ref{fig:LLs}) appear vertical between LL crossing points.
The sum $C_\text{tg} + C_\text{bg}$ is calibrated by graphically fitting the Landau fan at $D=0$, including LLs emerging from $|\nu_{12}|=4$.
We repeat this graphical fit at two extremes of the finite width $R_{xx}=0$ regions to generate upper and lower bound estimates for the capacitances and thus the extracted twist angles.
This leads to the error bounds $\theta_{12} = \SI{1.42\pm 0.07}{\degree}$ (see \extlbl~\ref{fig:fanfit}) and $\theta_{23}=\SI{-1.88\pm 0.08}{\degree}$.
The additional error in $\theta_{23}$ due to the finite width of the $R_{xx}$ peak is negligible.
As explained above, a small systematic error in $\theta_{23}$ is also possible (see Section~\ref{ssec:theta23}).

\subsubsection{Extracting layer-resolved constant density traces}\label{ssec:layer_resolved}
\begin{figure*}
    \centering
    \includegraphics[width=6.75in]{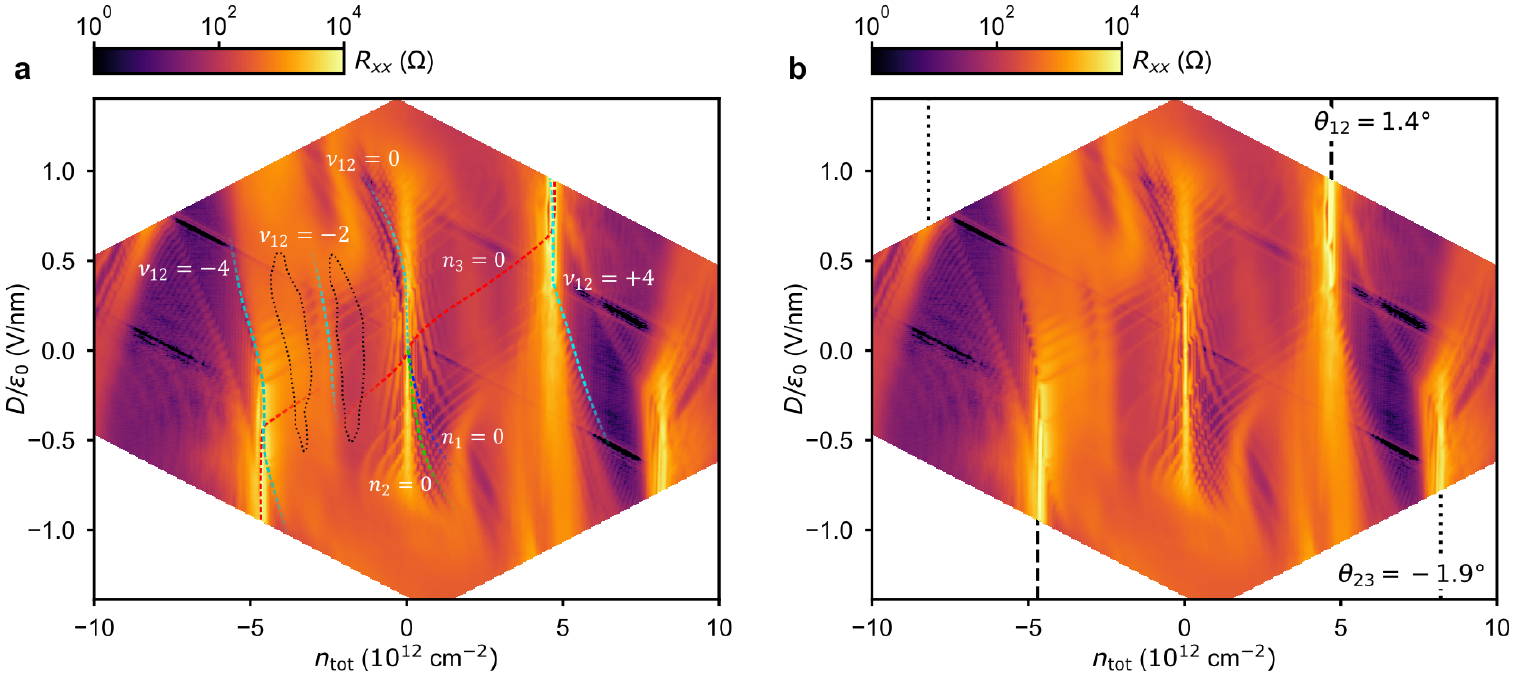}
    \caption{\figtitle{Layer-resolved constant density traces.}
    \panel{a} $R_{xx}$ versus $n_\text{tot}$ and $D$ at $B_\perp=\SI{1}{T}$ with important estimated traces as guide to the eye, indicated by labels.
    The dashed red line traces approximate charge neutrality of layer 3, $n_3 = 0$, indicated by the $N=0$ LL of layer 3.
    A resistance peak appears at the crossing point of $n_3 = 0$ and full filling of the pairwise \moire of layers 1 and 2, $\nu_{12} = 4$.
    At this point $n_\text{tot} \approx n_1+n_2 = n_{\text{s},12} = \SI{4.7e12}{cm^{-2}}$ confirming our value of $\theta_{12} = \SI{1.42}{\degree}$ (see \panel{b}).
    Semi-transparent traces indicate less well-defined layer character in regions where the hybridization between the layers is more pronounced.
    Superconducting pockets (see Fig.~\ref{fig:SC}) are outlined by black dotted lines.
    \panel{b} Same as \panel{a}, without the superimposed traces.
    }
    \label{fig:CNP_traces}
\end{figure*}

We can trace charge neutrality of layer 3, $n_3 = 0$, in the $(n_\text{tot},D)$ space by tracing the $N=0$ LL of the Dirac cone associated with layer~3 (\extlbl~\ref{fig:CNP_traces}\panel{a}, dashed red).
Starting at $(n_\text{tot},D)=(0,0)$ where all three layers are approximately charge neutral, we follow layer 3 charge neutrality in the positive $n_\text{tot}$ until we reach the $R_{xx}$ maximum, induced by the gap at full filling of the pairwise \moire from layers 1 and 2.
At this point $n_\text{tot}=n_{\text{s},12}$.
The full filling $\nu_{12} = +4$ is indicated in cyan.
For $n_\text{tot}$ beyond this trace the LLs from the Dirac cone associated with layer~3 become approximately vertical, indicating that the contribution of layers 1 and 2 to the DOS there is small.
A similar situation occurs for negative doping at $\nu_{12} = -4$.
The independent charge neutrality traces of layers 1 and 2 are indicated in blue and green, respectively.
Layers 1 and 2 produce clear independent LL sets for $n_\text{tot}>0, D<0$ where we can approximately trace $n_1 = 0$ and $n_2 = 0$ (\extlbl~\ref{fig:CNP_traces}\panel{a}, blue and green, respectively).
In the opposite quadrant, $n_\text{tot}<0, D>0$ it is harder to resolve the two sets and we therefore mark our best estimate for the combined $\nu_{12}=0$ trace.
The $\nu_{12} = -2$ trace is approximately half way between $\nu_{12} = 0$ and $\nu_{12} = -4$ and is important for the analysis of the flavor-symmetry-breaking phase transition and adjacent superconductivity.
This filling becomes less well-defined where it intersects with the $N=0$ LL of layer 3, as indicated by the strong downturn of this LL as it approaches the intersection.
The downturn indicates strong hybridization of the layer~3 Dirac cone with the flat quasi-band of layers 1 and 2 in agreement with our SF calculations (Fig.~\ref{fig:bands}\panel{d}).

\subsection{Phenomenological model}\label{ssec:phenomenological}
Our low energy phenomenological model consists of three Dirac cones with velocities $v_i$, for layer $i\in\{1,2,3\}$.
We allow linear shifts in energy with $D$ field, $\alpha_i = \partial \varepsilon_i/ \partial D$.
Focusing on $D=0$, the two velocity ratios, $v_1/v_3$ and $v_2/v_3$ fix the LL ordering between the three Dirac cones.
We use $R_{xx}(n,D)$ data taken at an intermediate field $B_\perp=\SI{1}{T}$ which is high enough so that the LLs are clearly visible but not too high, to ensure enough LLs fall within the density range in which they are visible.
At $B_\perp=\SI{1}{T}$ the $N=1$ LL from the fast Dirac cone (layer~3) appears at a high carrier density, $n_\text{tot}\approx \SI{2e12}{cm^{-2}}$, which corresponds to a large filling of the flat quasi-band, $\nu_{12}\approx1.7$.
At this filling the slow cones deviate substantially from linear dispersion due to hybridization, as is self-consistently confirmed by the SF calculations.
We therefore allow a quadratic term in the phenomenological dispersion, $\varepsilon_i(k)=\hbar v_i k + \hbar \beta_i k^2/2$.
The velocities are modified by the quadratic term to $v_i = v_i^0 + \beta_i k$.
Using the Bohr-Sommerfeld quantization condition, $\pi k_N^2l_\text{B}^2=2\pi(N+1/2+\gamma/2\pi)$, we get the modified LL energies, $\varepsilon_{N,i} = \operatorname{sgn}(N) v_i\sqrt{2\hbar eB|N|}+\beta_i eBN$.
Here, $\gamma=\pi$ is the Berry phase around the $K$ point of graphene.
The total number of parameters in our phenomenological model for $D=0$ is therefore 5 (two velocity ratios and three dispersion curvatures $\beta_i$).
The signs of $\alpha_i$ allow us to associate each of the LL sequences with a specific layer (Fig.~\ref{fig:LLs}\panel{c}).

\subsection{Electronic structure model}\label{ssec:SF}
To model the bands for our quasicrystalline system we employ a momentum space formulation for the interlayer tunneling that is valid for arbitrary twist angles between the layers \cite{koshino2015interlayer,amorim2018electronic,zhu2020twistedtrilayer}.
The general form for an eigenstate in the presence of interlayer tunneling is
\begin{equation}
\begin{split}
\ket{\psi_{\mathbf{k},n}}=&\sum_{\mathbf{G}_{2},\mathbf{G}_{3},\alpha} \phi^{n}_{1,\mathbf{k},\alpha}(\mathbf{G}_2,\mathbf{G}_3)\ket{1,\mathbf{k}+\mathbf{G}_2+\mathbf{G}_3,\alpha}\\
&+\sum_{\mathbf{G}_{1},\mathbf{G}_{3},\alpha} \phi^{n}_{2,\mathbf{k},\alpha}(\mathbf{G}_1,\mathbf{G}_3)\ket{2,\mathbf{k}+\mathbf{G}_1+\mathbf{G}_3,\alpha}\\
&+\sum_{\mathbf{G}_{1},\mathbf{G}_{2},\alpha} \phi^{n}_{3,\mathbf{k},\alpha}(\mathbf{G}_1,\mathbf{G}_2)\ket{3,\mathbf{k}+\mathbf{G}_1+\mathbf{G}_2,\alpha}
\end{split}
\end{equation}
where $\ket{\ell,\mathbf{k},\alpha}$ is the Bloch state on layer $\ell$ at momentum $\mathbf{k}$, $n$ is the eigenstate index, $\alpha=A,B$ is the sublattice index (spin degeneracy is implied throughout),  $\mathbf{G}_{\ell}$ are reciprocal lattice vectors for the layer $\ell$, and $\phi_{\ell,\mathbf{k},\alpha}^n$ are complex numbers.
The Bloch states satisfy $\ket{\ell,\mathbf{k}+\mathbf{G}_\ell,\alpha} = \ket{\ell,\mathbf{k},\alpha}$.
It is assumed in this formulation that the twist angles are incommensurate, i.e. there is no combination of non-zero $\mathbf{G}_{\ell},\mathbf{G}_{\ell^\prime}$, such that $\mathbf{G}_{\ell}+\mathbf{G}_{\ell^\prime}=0$.
The Hamiltonian consists of the intralayer part, $H_{\ell}(\mathbf{k})=\sum_{\alpha,\beta}h^\ell_{\alpha\beta}(\mathbf{k})\ket{\ell,\mathbf{k},\alpha}\bra{\ell,\mathbf{k},\beta}$, and the interlayer part, $H_{\ell\ell^\prime}(\mathbf{k})=\sum_{\mathbf{G}_1,\mathbf{G}_2,\mathbf{G}_3,\alpha,\beta} t(\mathbf{k}+\mathbf{G}_{123})e^{i\mathbf{G}_\ell\cdot\bm{\tau}^{\ell}_\alpha-\mathbf{G}_{\ell^\prime}\cdot\bm{\tau}^{\ell^\prime}_{\beta}}\ket{\ell,\mathbf{k}+\mathbf{G}_{123},\alpha}\bra{\ell^\prime,\mathbf{k}+\mathbf{G}_{123},\beta}$, where $\mathbf{G}_{123}\equiv\mathbf{G}_1+\mathbf{G}_2+\mathbf{G}_3$.
The full Hamiltonian is $H=\sum_{\mathbf{k},\ell} H_{\ell}(\mathbf{k}) +\sum_{\mathbf{k}} \left(H_{12}(\mathbf{k})+H_{23}(\mathbf{k}) + \mathrm{h.c.}\right)$.

We model the intralayer term as a nearest neighbor tight binding model of $p_z$ orbitals, ${h}_{BA}^{\ell}{(\mathbf{k})}=-t\sum_{i=1,2,3}e^{-i\mathbf{k}\cdot\mathcal{R}_{2\pi i/3}(\mathbf{\tau}^{\ell}_B-\tau^{\ell}_A)}=h_{AB}^{\ell}(\mathbf{k})^*$ and $h_{\alpha\alpha}^{\ell}(\mathbf{k})=0$, where $\mathcal{R}_{\theta}$ is a counter-clockwise rotation matrix by angle $\theta$, and $\tau^{\ell}_A=(0,0), \tau^{\ell}_B=\mathcal{R}_{\theta_\ell}(0,a_{cc})$ [$a_{cc}=\SI{1.43}{\AA}$] are the location of the two sublattices within the rotated graphene unit cell.
Note that this convention is $\SI{90}{\degree}$ rotated from the illustrations in the main text.
In the following, we define the $K$ point to be $\mathbf{K}=(K,0)$, and $\mathbf{K}^\ell=\mathcal{R}_{\theta_\ell}\mathbf{K}$, where $K=4\pi/(3\sqrt{3}a_{cc})$.
Here, $\theta_\ell$ is the counterclockwise rotation of the layer $\ell$, which we take to be $[\SI{1.4}{\degree},0,\SI{1.9}{\degree}]$.
The interlayer tunneling term is taken to be  the Fourier transform of the real space interlayer tunneling amplitude, $t(\mathbf{k})=\frac{1}{A_\mathrm{uc}}\int d^2 \mathbf{r} e^{-i \mathbf{k}\cdot\mathbf{r}}t(\mathbf{r})$, where $\mathbf{r}$ is the in-plane distance. $t(\mathbf{r})$ takes the Slater-Koster form \cite{koshino2015interlayer,slater1954simplified}
$ t(\mathbf{r}) = -t e^{-(R-a_{cc})/r_0} \frac{r^2}{R^2} + t_{\perp} e^{(R-d)/r_0} \frac{d^2}{R^2} $,
where $r=|\mathbf{r}|$, $R=\sqrt{r^2+d^2}$, $d=\SI{3.35}{\AA}$, $r_0=\SI{0.453}{\AA}$.

To produce quantitative agreement with experiment (Figs.~\ref{fig:LLs}\panel{b,c}), we use $t=\SI{3.1}{eV}$ (corresponding to a monolayer Dirac velocity of $v_\text{F}^{\text{MLG}} = \SI{1e6}{m/s}$) \cite{elias2011dirac,sokolik2017manybody}, and $t_{\perp}=\SI{0.43}{eV}$ (corresponding to an interlayer tunneling at $K$ of $t(\mathbf{K})=\SI{100}{meV}$).
We obtain these values by first using a perturbative approach (Methods~\ref{ssec:perturbative}) to get a coarse estimate and then fine-tuning them so that the microscopic model reproduces the Fermi velocity ratios extracted from the phenomenological model.
Fitting the resulting SF around each Dirac node yields effective Fermi velocities $(v_{1},v_{2},v_{3}) = (0.2,0.08,0.4)v_\text{F}^\text{MLG}$, consistent within 2\% with the velocity ratios extracted from Fig.~\ref{fig:LLs}\panel{c}.
Our value of the ratio $t(\mathbf{K})/v_\mathrm{F}$ is lower by $\approx 28\%$ from the one used in Ref.~\citenum{bistrizter2011moire}, and in agreement with recent experimental results \cite{turkel2022orderly}. 
We ascribe this difference to the reduced Fermi velocity in magic-angle TBG due to highly effective self-screening \cite{stauber2017interacting} induced by the high density of states of the flat bands.
To solve $H$, we construct the Hamiltonian matrix for a given $\mathbf{k}$ using the set of basis states  $\left\{\ket{1,\mathbf{k}+\mathbf{G}_2+\mathbf{G}_3,\alpha},\ket{2,\mathbf{k}+\mathbf{G}_1+\mathbf{G}_3,\alpha},\ket{3,\mathbf{k}+\mathbf{G}_1+\mathbf{G}_2,\alpha}\right\}$, for all $\mathbf{G}_{\ell}$ up to a cutoff $|\mathbf{G}_{\ell}| < \Lambda$.
We use the cutoff $\Lambda=7.1K$ for the SF calculations.
We take two further approximations: the first-harmonic approximation is made by only keeping interlayer tunneling terms with $t(\mathbf{k}+\mathbf{G}_{123})$ for which $|\mathbf{k}+\mathbf{G}_{123}|<1.4 K$.
Second, since only $t(\mathbf{k})$ for $|\mathbf{k}|\approx K$ enter, we expand $t(\mathbf{k})\approx t_0+t_1(|\mathbf{k}|-K) + t_2(|\mathbf{k}|-K)^2$, providing a computational simplification [for our parameters, $t_0=\SI{0.10}{eV}$, $t_1=\SI{-0.20}{eV \AA}$, $t_2=\SI{0.33}{eV \AA^{2}}$].
The Hamiltonian is sparse and the states closest to $E=0$ can be obtained numerically, for example, via the shift-invert Lanczos algorithm.
We obtain the eigenvalue $\varepsilon_{\mathbf{k} n}$ and eigenvectors $\ket{\psi_{\mathbf{k} n}}$, from which we can extract the layer-resolved spectral weights $w_{n\mathbf{k}\ell} = \sum_{\alpha}|\phi^n_{\ell \mathbf{k} \alpha}(0,0)|^2$.
We compute the $300$ eigenvalues closest to zero, which is sufficient to obtain all eigenvalues within an energy window of $\pm\SI{0.2}{eV}$.

The SF shown in Fig.~\ref{fig:bands} is obtained by plotting $\varepsilon_{\mathbf{k} n}$ along a path in momentum space, with the width of the line proportional the total spectral weight $\sum_\ell w_{n\mathbf{k}\ell}$, and the color indicating the direction of the vector $(w_{n\mathbf{k}1},w_{n\mathbf{k}2},w_{n\mathbf{k}3})$.

Selected constant-energy cuts of the SF are shown in \supplbl~\ref{fig:SF-2D-cuts} and the full sequence of energy cuts is given in Supplementary Video 2.

\begin{figure*}
    \centering
\includegraphics[width=180mm]{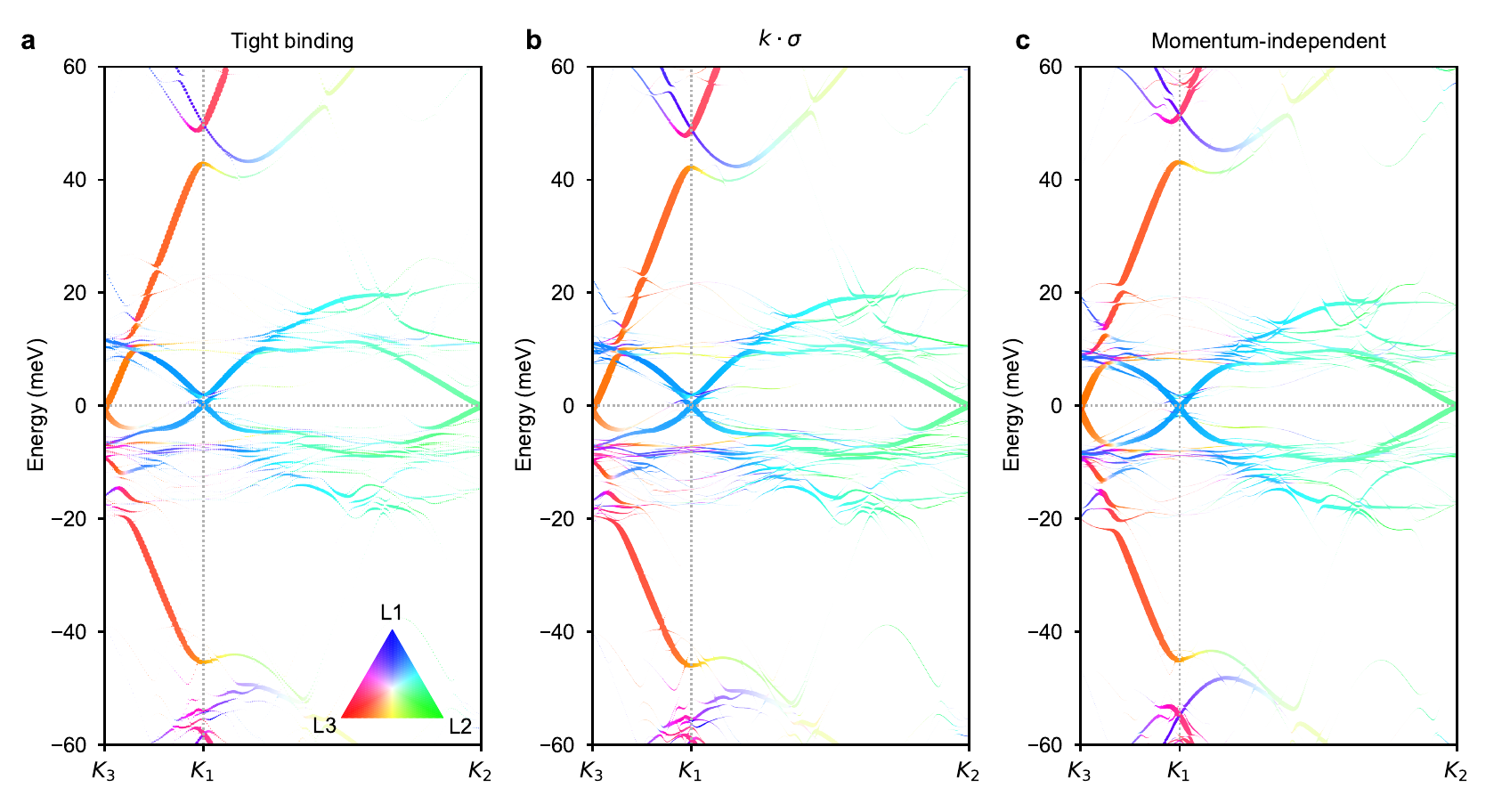}
\caption{\figtitle{Comparison of spectral function approximations.}
The spectral function calculated for \panel{a} the tight binding intralayer dispersion, \panel{b} the effective $k\cdot\sigma$ Dirac cone dispersion, and \panel{c} momentum independent interlayer tunneling (keeping only $t_0$ in the expansion of $t(\mathbf{k})$), for the $k\cdot\sigma$ dispersion. The three methods show excellent agreement.
}
    \label{fig:kdsvsTB}
\end{figure*}

We also verify that replacing the tight binding intralayer Hamiltonian with the effective $k\cdot\sigma$ Dirac cones, 
$h_{AB}^{\ell}(\mathbf{k})=v_\text{F} [(k_x-K_x^\ell)-i (k_y-K_y^\ell)]e^{i\theta_\ell}$, does not lead to any significant changes.
This expansion is valid as all momenta involved lie near the $\mathbf{K}^\ell$ points of each layer.
\extlbl~\ref{fig:kdsvsTB}\panel{a,b} show the computed SF for the model with the tight binding and effective Dirac cone, respectively, showing excellent agreement.
In \extlbl~\ref{fig:kdsvsTB}\panel{c}, we show the effect of taking a momentum-independent first-harmonic approximation for the interlayer hopping $t(\mathbf{K})\approx t_0 = \SI{0.1}{eV}$.
This final version of the theory is the trilayer generalization of the Bistritzer-MacDonald continuum model~\cite{bistrizter2011moire} which is fully characterized by the twist angles, Fermi velocity, and interlayer tunneling strength.
In this limit, the lattice scale is forgotten and only the moir\'e scales remain, exemplifying the name ``moir\'e quasicrystal''.
We remark that although we have formulated our theory in terms of the monolayer reciprocal lattice vectors, we could have equally well formulate this theory using only the moir\'e reciprocal lattice vectors, as in the Bistritzer-MacDonald model.

\subsection{DOS calculation}\label{ssec:DOS}
The density of states is obtained by performing the calculation outlined in Sec.~\ref{ssec:SF} for a dense grid of $\mathbf{k}$ about the K points.
We define the spectral function as $A_\ell(\omega,\mathbf{k})=\sum_{n} w_{n\mathbf{k}\ell}\delta(\omega-\varepsilon_{n\mathbf{k}})$. 
In practice, we replace the $\delta$-function with a Lorentzian $ L_\sigma(x)=\frac{\sigma/\pi}{x^2+\sigma^2}$ with broadening $\sigma=0.5$meV.
The layer-resolved DOS is then defined as $D_\ell(E)=4\int_{K} \frac{d^2\mathbf{k}}{4\pi^2} A_\ell(\omega,k)$, where the integral is within a small window of the $K$ points, and the factor of 4 is to account for the spin and valley degeneracy.
The total DOS is simply $D(E)=\sum_\ell D_\ell(E)$.
We sample $\mathbf{k}$ in a grid of $250\times 250$ points within a \SI{0.3}{\AA^{-1}}$\times$ \SI{0.3}{\AA^{-1}} square about the $K$ point.
We use a cutoff of $\Lambda=4.1K$ for the DOS calculation and solve for the $110$ eigenstates closest to $E=0$, sufficient to accurately obtain the spectral function for $|E|<\SI{0.2}{eV}$.

\begin{figure*}
    \centering
\includegraphics[width=180mm]{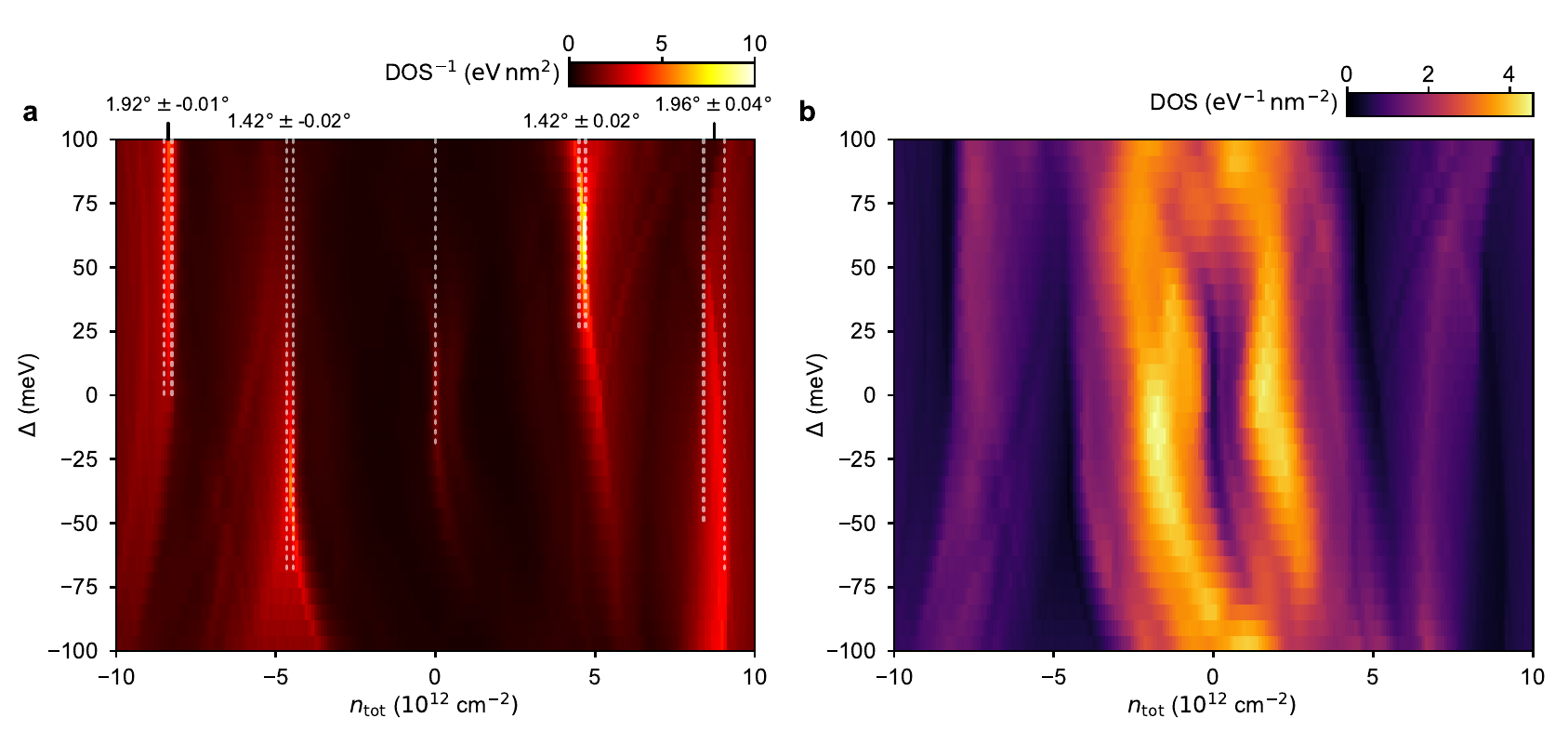}
\caption{\figtitle{Calculated density of states maps.}
\panel{a} Inverse density of states (DOS) as a function of density and layer potential, calculated for $(\theta_{12},\theta_{23}) = (\SI{1.4}{\degree},\SI{-1.9}{\degree})$.
Shown are the pairwise \moire angles inferred from the inverse DOS peaks, in excellent agreement with the angles input to the calculation, validating the procedure of extracting the twist angles from the magnetotransport data.
\panel{b} Density of states as a function of density and layer potential.
The high density of states region in $n_\text{tot}<0$ generally overlaps with the right superconducting pocket.}
    \label{fig:idos}
\end{figure*}

To compare with experiment, we also compute the DOS as a function of density and layer potential $\Delta$.
The total density at a given energy can be obtained by $N(E)=4\int\frac{d^2\mathbf{k}}{4\pi^2}\sum_{n\ell}w_{n\mathbf{k}\ell}\Theta(E-\varepsilon_{n\mathbf{k}})$, where $\Theta(x)$ is the Heaviside step function.
Since we only obtain a fixed number of eigenvalues, there is an overall offset in the density calculated in this way which can in principle vary as a function of displacement field, even with all other parameters kept constant.
In the plot of DOS as a function of density, we assume that the density offset is constant over all $\Delta$ while keeping all other parameters fixed, and set the density at the Dirac cones at $\Delta=0$ to be $N=0$.
We have verified that this assumption is quite accurate within the range of interest by comparison with the DOS obtained for a commensurate approximation for which the charge neutrality point is known.
\extlbl~\ref{fig:idos}\panel{a} shows the inverse total DOS as a function of density and layer potential.
The peaks in the inverse DOS correspond to more incompressible regions, and their locations are in excellent correspondence with the experimentally observed high resistance regions.

\subsection{Perturbative estimation of hopping energies $\omega/v_\text F$}\label{ssec:perturbative}
\begin{figure}
    \centering
\includegraphics[width=88mm]{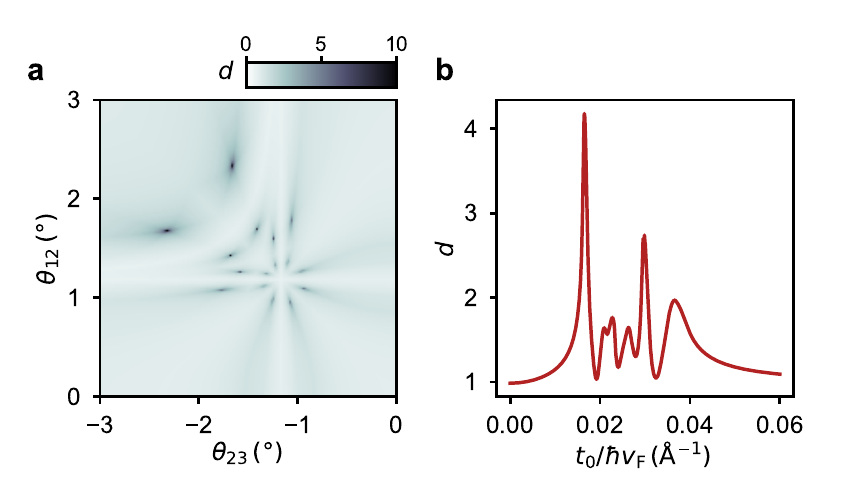}
\caption{\figtitle{Matching function for twist angles and model parameters.}
\panel{a} Matching function $d$ from Eq.~\eqref{eq:cost} plotted for the target velocities $(0.2,0.51,1)$ with $t_0/\hbar v_\text F = \SI{0.02}{\AA^{-1}}$.
\panel{b} Matching function versus $t_0/\hbar v_\text F$ assuming the same target velocities, averaged over 100 points in the range $\SI{1.35}{\degree} < \theta_{12} < \SI{1.45}{\degree}$, $\SI{-1.95}{\degree}< \theta_{23} < \SI{-1.85}{\degree}$. The dominant peak is at $\SI{0.016} {\AA^{-1}}$.}
    \label{fig:cost}
\end{figure}

The renormalized Dirac cone velocity of layer~2 in TTG has the following expression to first-order in interlayer tunneling \cite{zhu2020twistedtrilayer}: 
\begin{equation}
     v_2^* = \frac{1-3(\alpha_{12}^2+ \alpha_{23}^2)}{1+6(\alpha_{12}^2+\alpha_{23}^2)}v_\text F
\end{equation}
where $\alpha_{ij} = t_0/ (\hbar v_\text F k_{\theta_{ij}})$, $k_{\theta} = 8\pi \sin(\theta/2)/(3a)$, assuming the interlayer AA/BB and AB/BA hopping elements each take the value $t_0$.
If we ignore tunneling between layer~1 and layer~3, their velocities are obtained by setting $\alpha_{23} =0$ or $\alpha_{12} =0$, respectively.
This provides us an approximate mapping from the three twist angles to the three velocities.
The reverse mapping, from velocities to twist angles, is not necessarily unique.
To see this, we plot a matching function, $d$, showing the deviation of the velocities $(v_1^*,v_2^*,v_3^*)$ from a target set of velocities $(v_1^\text T, v_2^\text T, v_3^\text T)$ in \extlbl~\ref{fig:cost}.
The matching function is defined as 
\begin{equation}\label{eq:cost}
    d = \sqrt{\frac{\text{mean}(\mathbf{v})}{\max(\mathbf{v})- \min(\mathbf{v})}},\qquad \mathbf{v} = \left(\frac{v_1^\text T}{ v_1^*}, \frac{v_2^\text T}{v_2^*}, \frac{v_3^\text T}{ v_3^*}\right)
\end{equation}
where we assume $v_1^* \leq v_2^* \leq v_3^*$ and similarly for the target velocities, $v_i^\text{T}$.
The matching function diverges at an exact solution, which corresponds to $\mathbf{v} = (1,1,1)$.
From \extlbl~\ref{fig:cost}\panel{a}, we see that there can be a discrete set of possible solutions in twist angle space for a generic set of target velocities.

Given a set of known twist angles and velocities, the perturbative formulas can also be used to approximate $t_0/\hbar v_\text F$ in the TTG sample.
We do this near the measured twist angles $(\theta_{12},\theta_{23}) = (\SI{1.4}{\degree},\SI{-1.9}{\degree})$ and normalized velocities $(0.2,0.51,1)$ and find a closest fit of $t_0/ \hbar v_\text F \approx \SI{0.016} {\AA^{-1}}$ (\extlbl~\ref{fig:cost}\panel{b}).
We therefore used $v_\text F = \SI{1e6}{m/s}$ and $t_0=\SI{0.105} {eV}$ as a starting point for our SF calculation.

\subsection{Estimating layer potentials}\label{ssec:electrostatics}
\begin{figure*}
    \centering
    \includegraphics[width=180mm]{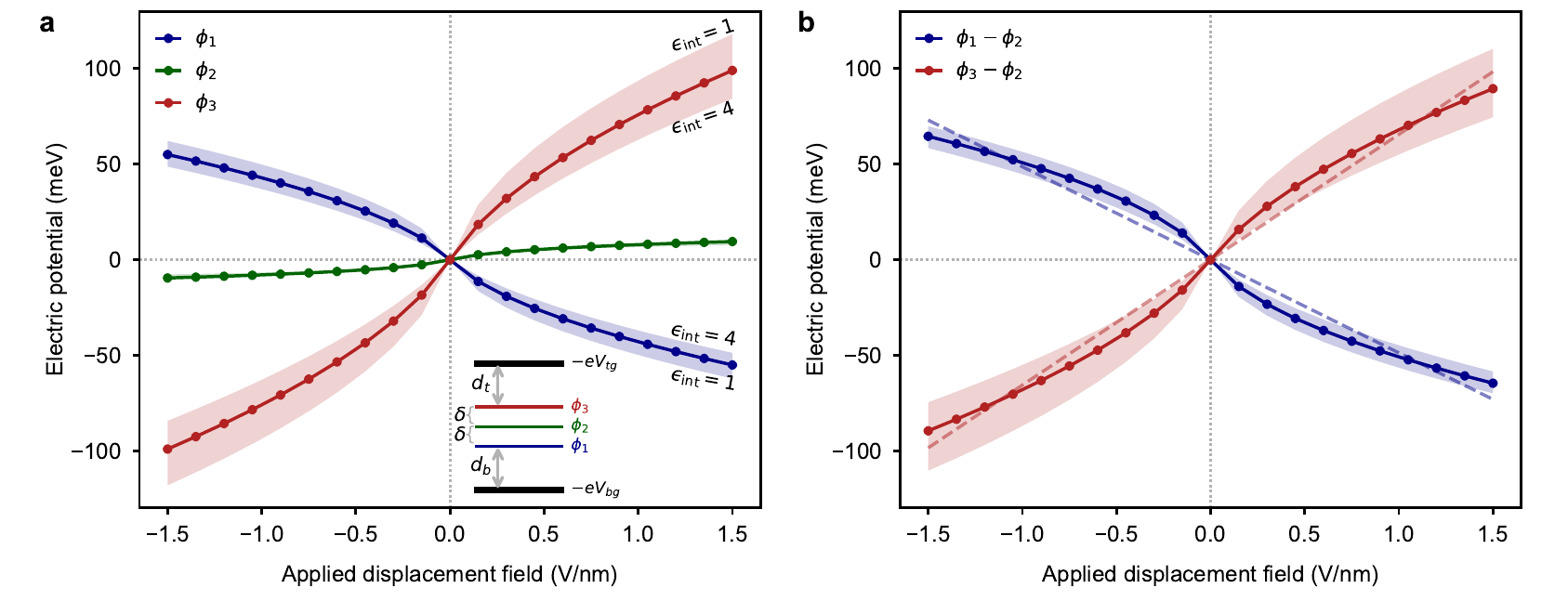}
    \caption{\figtitle{Electrostatic layer potential calculation.}
    \panel{a} Calculated electric potential energies $\phi_i$ (for layer index, $i$) as a function of externally applied electrical displacement field $D/\epsilon_0$ for $n_\text{tot}=0$ using Fermi velocities extracted from the spectral function calculation; used to estimate correspondence between layer potential imbalance $\Delta$ and $D/\epsilon_0$ in the main text.
    Solid lines are shown for $\epsilon_\text{int}=2.5$, while the shaded regions indicate the extreme cases described in Methods~\ref{ssec:electrostatics}.
    \panel{b} Potential energy difference $\phi_i - \phi_2$ between layers $i$ and 2, since to first order the relative energy shifts between the layers is the relevant quantity.
    Linear fits (dashed lines) illustrate the slight non-linear deviations.
    }
    \label{fig:electrostatics}
\end{figure*}

To map the experimentally applied $D$ field to the layer potentials $\Delta_i$ for $i=\lbrace1,2,3\rbrace$ applied in the SF calculation, we perform a charge balance calculation to simulate the electrostatics of the experimental device.
We assume linear Dirac dispersions for the three layers with Fermi velocities $(v_{1},v_{2},v_{3}) = (0.2,0.08,0.4)v_\text{F}^\text{MLG}$, $v_\text{F}^\text{MLG} = \SI{1e6}{m/s}$ extracted from the SF calculations and the phenomenological model.
Each Dirac cone is assumed to reside on the $z$-plane of one of the layers, separated by the graphene interlayer distance we take to be $\delta =\SI{0.33}{nm}$ (\extlbl~\ref{fig:electrostatics}\panel{a}, inset).
The external $D$ field is simulated by including gate voltages $V_\text{tg}$ and $V_\text{bg}$ at distances $d_\text{t}$ and $d_\text{b}$, calibrated from the experiment, respectively.
The system is modeled as a 5-parallel-plate capacitor with the graphene layers fixed to ground while the gate voltages vary.
The electric potential energies, $\phi_i$ on the three graphene layers lead to layer charge densities,
\begin{align}\label{eq:charge}
    -en_1 &= C_\text{bg} ( V_\text{bg} - \phi_1/e ) + C_0 ( \phi_2 - \phi_1 ) /e \nonumber \\ 
    -en_2 &= C_0 ( \phi_1 - 2\phi_2 + \phi_3 ) /e \\
    -en_3 &= C_\text{tg} ( V_\text{tg} - \phi_3/e ) + C_0 ( \phi_2 - \phi_3 ) /e \nonumber 
\end{align}
where the electron densities are given by the integrated DOS up to the Fermi energy ($\mu_i = -\phi_i$ for grounded layers) for monolayer graphene, $n_i = -\frac{\operatorname{sgn}(\phi_i)}{\pi}\pqty{\frac{\phi_i}{\hbar v_i}}^2$.
Here, $C_\text{tg}$ and $C_\text{bg}$ are the geometric top and bottom capacitances per unit area used throughout the paper, and $C_0 = \epsilon_\text{int}\epsilon_0 / \delta$ is the capacitance between graphene layers.
Since the interlayer dielectric constant $\epsilon_\text{int}$ for two-angle TTG is not known, we compute the results for a range of values between 1 and 4 (the vacuum limit and close to the value extracted from Bernal bilayer graphene in large magnetic fields \cite{zibrov2017tunableinteracting}, respectively), and use these extremes to provide an estimate of the error (shaded regions in \extlbl~\ref{fig:electrostatics}).
The charge-balance Eqs.~\ref{eq:charge} are solved self-consistently for a set of gate voltages (chosen to vary $D/\epsilon_0$, keeping $n_\text{tot}=0$) to determine the layer potential energies $\phi_i$ in equilibrium.
The potential imbalance $\Delta$ employed in the SF calculations are then selected to roughly correspond to the self-consistent values of $\phi_i$ determined in the electrostatics simulation.
The slight non-linearity of $\phi_i(D)$ and the asymmetry $\phi_1 - \phi_2 \neq -(\phi_3 - \phi_2)$ (\extlbl~\ref{fig:electrostatics}\panel{b}) are ignored in the SF calculation, where the goal is simply to observe the zeroth-order effect of a potential imbalance across the layers.

\subsection{Symmetry-breaking phase transition analysis}\label{ssec:DR}
\begin{figure*}
    \centering
    \includegraphics[width=1\linewidth]{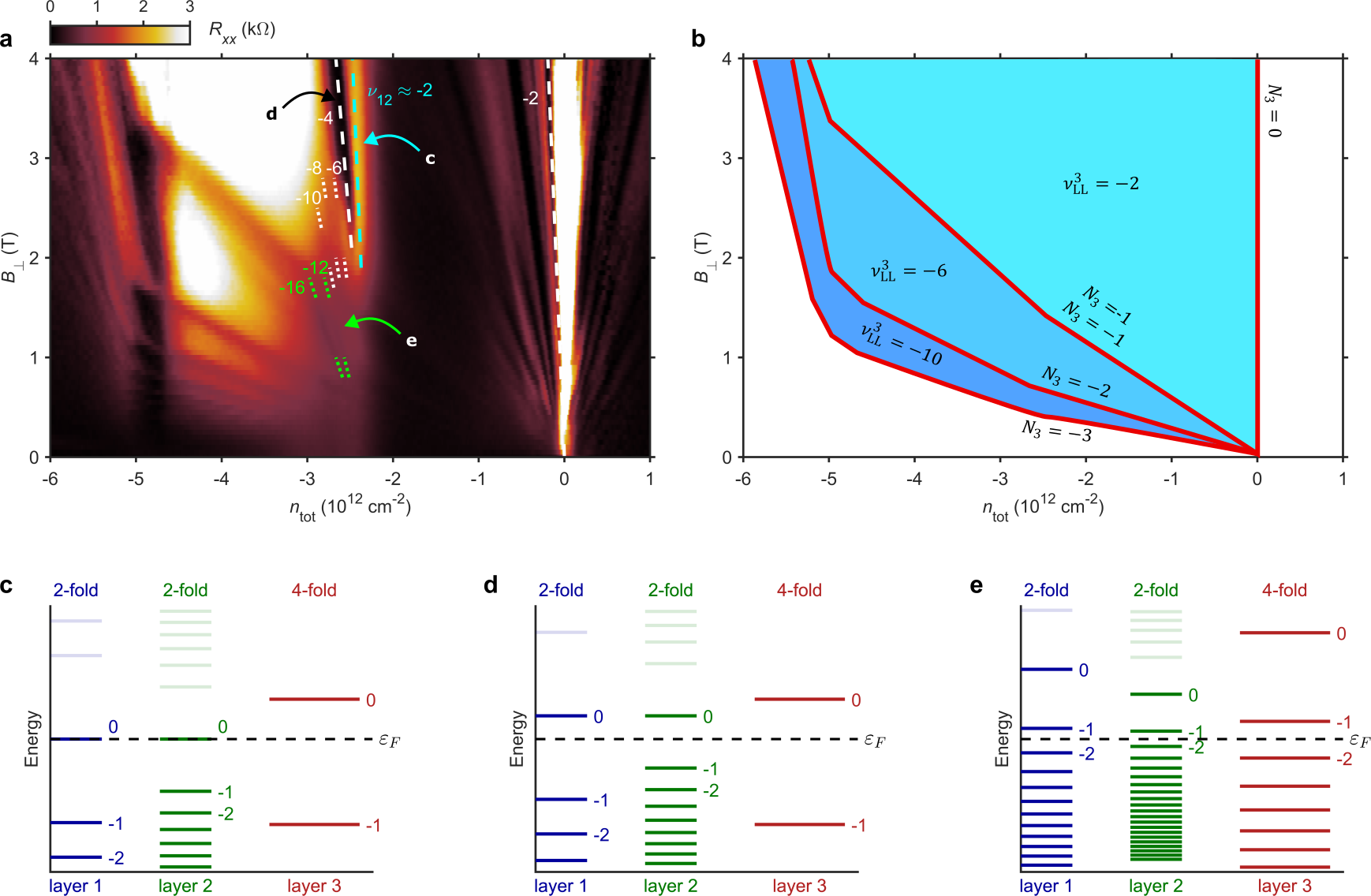}
    \caption{
    \figtitle{Flavor-symmetry-breaking phase transition.}
    \panel{a} $R_{xx}$ versus $n_\text{tot}$ and $B_\perp$ taken at $T=\SI{500}{mK}$, showing a Landau fan emerging from \moire filling $\nu_{12} \approx -2$ indicating a flavor-symmetry-breaking phase transition near that carrier density.
    \panel{b} Illustration of the layer~3 LLs and their contributions to the Chern number, $\nu_\text{LL}^3$, in the gaps between the LLs.
    \panel{c} The situation at $\nu_{12} \approx -2$ (\panel{a}, dashed cyan).
    The Fermi energy is in the $N=0$ LLs of layers 1 and 2, and in the gap between LL $N_3 = 0$ and $N_3 = -1$ of layer 3.
    Semi-transparent LLs indicate inaccessible levels before the phase transition ($\nu_{12} > -2$).
    \panel{d} In the first LL gap beyond the phase transition (\panel{a}, dashed white), the Fermi level is in the gap between LLs $0$ and $-1$ in all three layers.
    Layer 3 contributes $C=-2$ while layers 1 and 2 each contributes $C=-1$ due to their reduced degeneracy, accounting for the observed slope $C=-4$.
    \panel{e} In the $C=-12$ gap (\panel{a}, dotted green), the Fermi energy is between LLs $-1$ and $-2$ in all three layers.
    The Chern contributions are $-1,-2=-3$ for layers 1 and 2, and $-2-4=-6$ for layer 3, in total $-3-3-6 = -12$.
    }
    \label{fig:revival}
\end{figure*}

\extlbl~\ref{fig:revival}\panel{a} shows the Landau fan originating from \moire filling $\nu_{12} \approx -2$ due to the phase transition near that filling.
Beyond this symmetry-breaking phase transition, layers 1 and 2 become flavor-polarized \cite{wong2020cascade,zondiner2020cascade}.
The data can be understood by attributing the phase transition to the states associated mostly with layers 1 and 2, acting as a flat quasi-band, whereas the Dirac cone of layer 3 remains a spectator.
Tuning the carrier density to $\nu_{12} \approx -2$, two of the flavors are emptied and are reset to the $N=0$ LL.
The new sequences of LLs from layers 1 and 2 become 2-fold degenerate (\extlbl~\ref{fig:revival}\panel{c}).
The cyan dashed line in \extlbl~\ref{fig:revival}\panel{a} traces the center of the $N=0$ LLs of layers 1 and 2, indicating $\nu_{12} \approx -2$.
Along this line, the Fermi energy is in the gap between the $N=0$ and $N=-1$ LLs of layer 3.
\extlbl~\ref{fig:revival}\panel{b} shows the traces of the different LLs of layer 3, as well as its LL filling fractions $\nu^3_\text{LL}$.
The cyan dashed line has a slope $C = \phi_0 \partial n_{12} / \partial B = -2$.
Since the zeroth LL of layer 3 is 4-fold degenerate, it contributes $C=-2$.
Subtracting this contribution reveals that the two occupied flavors of the flat quasi-band contribute together $C=0$.
We note that the absence of gaps in our system \cite{mora2019flatbands} implies that there is always finite DOS from the occupied flavors in the broken-symmetry phase.

The following minimum in $R_{xx}$ has a total slope $C=-4$ (\extlbl~\ref{fig:revival}\panel{a}, dashed white), consistent with a contribution $C=-2$ from the $N=0$ LL of layer~3 and another $C=-1$ from each of the $N=0$ LLs of the two reset flavors of layers 1 and 2 (\extlbl~\ref{fig:revival}\panel{d}).
We also observe two additional weak $R_{xx}$ minima with slopes $C=-12$ and $C=-16$ (\extlbl~\ref{fig:revival}\panel{a}, dotted green), at the $\nu_\text{LL}^3 = -6$ gap between LLs $N=-1$ and $N=-2$ of layer~3 (\extlbl~\ref{fig:revival}\panel{b}).
We interpret the $C=-12$ gap as illustrated in \extlbl~\ref{fig:revival}\panel{e}.
It has a contribution of $C=-6$ from layer~3 and $C=-1-2=-3$ from LLs $N=0$ and $N=-1$ of each of the two reset flavors of layers 1 and 2.
Similarly, the $C=-16$ can be interpreted as including an extra $C=-2$ from each of the two emptied cones.

The data were taken at $T=\SI{500}{mK}$ between voltage contacts 4 and 5 (\figlbl~\ref{fig:setup}\panel{e}, inset), where $\theta_{12} \approx \SI{1.4}{\degree}$, similar to that of the main contact pair.

\subsection{Analyzing the density per flavor in the superconducting pockets}\label{ssec:SC-analysis}
The superconducting state comprises two pockets, between $-2\lesssim\nu_{12}\lesssim-1.3$ and $-3.3 \lesssim \nu_{12} \lesssim -2.7$, respectively, both bounded by $|D/\epsilon_0|\lesssim\SI{0.5}{V/nm}$ at $T=\SI{160}{mK}$ (Fig.~\ref{fig:SC}\panel{a}).
It is instructional to consider the filling per flavor of the lower filling (active) flavors, $\nu_\text{f} = \operatorname{sgn}(\nu)(|\nu|-(4-g))/g$, where $g>0$ is the (integer) number of active flavors.
Using this definition, the first (right, $g=4$) pocket appears between $-0.5\lesssim\nu_\text{f,12}\lesssim-0.325$ and the second (left, $g=2$) pocket between $-0.65\lesssim\nu_\text{f,12}\lesssim-0.35$, similarly to the first pocket, which is truncated at $\nu_\text{f,12}=-0.5$ by the phase transition.
This suggests that the left ($g=2$) superconducting pocket may be related to the right one.
Superconductivity in magic-angle TBG and magic-angle TTG seems to require a parent state with broken flavor symmetry, $g \neq 4$.
Our first pocket is hosted by a flavor-symmetric ($g=4$) state - different than the magic-angle scenario. 
Our second pocket, although hosted by a flavor-symmetry-broken state ($g=2$), could be a duplicate of the first pocket.
Interestingly, despite the approximately electron-hole-symmetric DOS obtained in our calculations (Fig.~\ref{fig:bands}\panel{b}), we do not observe superconductivity on the electron-doped side, $n_\text{tot}>0$.
However, the appearance of a symmetry-breaking transition only on the hole-doped side may indicate a larger maximal DOS there.

\subsection{Calculating separation of scales}\label{ssec:SOS}
Figure~\ref{fig:sos}\panel{h} shows the separation of length scales, $\gamma$, calculated for intermediate carrier densities $n_\text{tot} \gtrsim \SI{1e12}{cm^{-2}}$, such that $n_\text{tot} \sim \lambda_\text{m}^{-2}$.
The separation of scales distinguishes \moire periodic crystals ($\gamma \gg 1$) from MQCs ($\gamma \gtrsim 1$) at the densities in question.
Below we define $\gamma$ and describe its calculation.

We take monolayer graphene lattice vectors as the column vectors of $A_0 = (\begin{matrix} \bold{a}_1 & \bold{a}_2\end{matrix})$, with $\bold a_1 = a \big(\begin{smallmatrix} 1 \\ 0 \end{smallmatrix} \big)$, $\bold a_2 = a \big(\begin{smallmatrix} 1/2 \\ \sqrt{3}/2 \end{smallmatrix}\big)$, and $a$ the monolayer graphene lattice constant.
The generators of the reciprocal lattice of layer~2 are the column vectors of $G_0 = 2\pi (A_0^{-1})^\text{T}$, and those of layer~1 and 3 are given by the column vectors of $G_1 = R(-\theta_{12}) G_2$ and $G_3 = R(\theta_{23}) G_2$, respectively, where $R(\theta) = \big(\begin{smallmatrix} \cos{\theta} & -\sin{\theta} \\ \sin{\theta} & \cos{\theta} \end{smallmatrix} \big)$ is the rotation matrix.
The three pairwise \moire reciprocal lattice generators are the column vectors of $G_{ij} = G_i - G_j$, with $(i,j) = \lbrace(1,2), (2,3), (1,3)\rbrace$.
The pairwise \moire real space lattice vectors are the column vectors of $A_{ij}=2\pi (G_{ij}^{-1})^\text{T} =\big(\begin{matrix} \bold{a}_{ij}^1 & \bold{a}_{ij}^2 \end{matrix}\big)$, with a unit cell area $S_{ij}=|\bold{a}_{ij}^1 \times \bold{a}_{ij}^2|$ and a real space lattice constant $\lambda_{ij}=|\bold{a}_{ij}^1|$.
For a given set of twist angles $(\theta_{12},\theta_{23})$ we calculate $\lambda_{12}, \lambda_{23}$, and $\lambda_{13}$ and sort them such that $\lambda^{(1)} > \lambda^{(2)} > \lambda^{(3)}$.
Additionally, we keep only unique $\lambda$ values.

We define the separation of scales as the smallest (but greater than $1$) ratio of the different scales in the system, $\gamma = \min\{\lambda^{(1)}/\lambda^{(2)},\lambda^{(2)}/\lambda^{(3)},\lambda^{(3)}/a\}$, thus $\gamma$ takes values greater than 1.
For example, in mirror-symmetric TTG, $\lambda_{12} = \lambda_{23}$ and $\lambda_{13} \rightarrow \infty$, therefore the list of lengths is $\lambda^{(i)} = \{\lambda_{12}, a\}$, and the separation of scales is $\gamma = \lambda_{12}/a \gg 1$.
Mirror-symmetric TTG is described by the bright diagonal line in Fig.~\ref{fig:sos}\panel{h}.
magic-angle TTG appears on this line with $\theta_{12} = -\theta_{23} \approx \SI{1.5}{\degree}$ and $\gamma \approx 38$.
The horizontal line with $\theta_{23}=0$ corresponds to twisted mono-bilayer graphene and shows a large separation of scales, $\gamma \approx a/\theta_{12} \gg 1$. 
This line also describes the separation of scales of TBG with a twist angle $\theta = \theta_{12}$.

The examples given above are of \moire periodic crystals and thus they all have a large separation of scales, $\gamma \gg 1$.
However, most of the space of twist angles comprises non-mirror-symmetric TTG with two non-zero twist angles.
In non-mirror-symmetric TTG (two-angle TTG), three different \moire lattices form, at least two of which have comparable length scales.
They are generally incommensurate, as exemplified by real-space images in Refs.~\citenum{zuo2018scanning,huang2021imaging,li2022symmetrybreaking}.
Therefore, the separation of scales is low, $\gamma \gtrsim 1$, reflecting a \moire quasicrystal at the densities in question, $n_\text{tot} \sim \lambda_{ij}^{-2} \ll a^{-2}$.
For example, the star in Fig.~\ref{fig:sos}\panel{h} marks the system reported here, with pairwise \moire lattices constants $\lambda_{12} = \SI{9.9}{nm}$, $\lambda_{23}=\SI{7.5}{nm}$, and $\lambda_{13} = \SI{30.6}{nm}$, resulting in a low separation of scales, $\gamma = \lambda_{12}/\lambda_{23} = 1.3$.
We further validate that small $\gamma$ reflects MQCs by performing spectral function calculations for selected twist angles with $\gamma \gtrsim 1$ (\supplbl~\ref{fig:other-angles}), all showing quasiperiodic features in their SFs, in contrast to systems with $\gamma \gg 1$ such as magic-angle graphene.

For larger twist angles, $\theta \gtrsim \SI{5}{\degree}$, the carrier densities at full filling of the pairwise \moire lattices are substantially larger, $n_{\text{s},ij} \gtrsim \SI{5e13}{cm^{-2}}$.
In this situation, densities of $n_\text{tot} \sim \SI{1e12}{cm^{-2}}$ are much smaller than full filling, and the pairwise periodicity can approximately be ignored.
Therefore, we expect parts of the space of twist angles to be \moire-of-\moire periodic crystals even at densities as high as $n_\text{tot} \sim \SI{1e12}{cm^{-2}}$.
At these larger twist angles the thin orange strip in Fig.~\ref{fig:sos}\panel{d} spans a broader density range.
By limiting our discussion to $\theta_{ij} < \SI{3}{\degree}$, we guarantee that $\lambda_{ij}^{-2}$ are never much larger than the densities in question and the incommensurate pairwise \moire lattices cannot be ignored.

\subsection{DOS calculation under finite magnetic field}\label{ssec:SF_finite_B}
In this section, we describe the formalism used to calculate the DOS in a finite magnetic field.
We first construct a commensurate approximation (periodic approximant), in which there is an exact periodicity for the \moire patterns formed by $\theta_{12}$ and $\theta_{23}$ \cite{mora2019flatbands}.
We then apply a magnetic field to the trilayer Bistritzer-MacDonald model via minimal coupling and work in the LL basis of the monolayer Dirac cones.

We first define the matrix $\mathbf{A}_0 = a\begin{pmatrix}1&1/2\\0&\sqrt{3}/2\end{pmatrix}$ whose columns are the monolayer lattice vectors ($a= \SI{0.246}{nm}$).  
The reciprocal lattice vectors are then given by the columns of $\mathbf{G}_0 = 2\pi\mathbf{A}_0^{-T}$.
For the twisted trilayer, we have $\mathbf{G}_\ell = \mathcal{R}_{\theta_\ell} \mathbf{G}_0$ for each layer $\ell$.
The moir\'e reciprocal lattice vectors between layers $\ell$ and $\ell^\prime$ are then given by $\mathbf{G}_{\ell \ell^\prime} = \mathbf{G}_\ell - \mathbf{G}_{\ell^\prime}$ and moir\'e lattice vectors $\mathbf{A}_{\ell\ell^\prime}=2\pi\mathbf{G}_{\ell\ell^\prime}^{-T}$.

To construct a periodic approximant for a target system at twist angles $(\theta_1,\theta_2,\theta_3)$, we find integers $(n_1,n_2,n_3,n_4)$ such that 
\begin{equation}
n_1 \mathbf{a}^{12}_{1} + n_2 \mathbf{a}^{12}_{2} \approx n_3 \mathbf{a}^{23}_{1} + n_4 \mathbf{a}^{23}_{2}
\label{eq:commens}
\end{equation}
 where $(\mathbf{a}^{\ell\ell^\prime}_i)_j = (\mathbf{A}_{\ell\ell^\prime})_{ji}$ is the $i$th \moire lattice vector formed by $\ell,\ell^\prime$.
Then, we allow for slight deformations of the middle ($\ell=2$) layer in the form of a small rotation $\theta_2\rightarrow \theta_2+\delta\theta_2$ and scaling factor $s$, $\mathbf{A}_2\rightarrow s^{-1}\mathcal{R}_{\delta\theta_2}\mathbf{A}_2$, such that Eq.~\ref{eq:commens} is exactly satisfied.
The resulting system is exactly periodic with  supermoir\'e reciprocal lattice vectors
\begin{equation}
\mathbf{G}_{\mathrm{SM}}=\mathbf{G}_{12}\begin{pmatrix}n_1&n_2\\-n_2&n_1+n_2\end{pmatrix}^{-1}
= \mathbf{G}_{23}\begin{pmatrix}n_3&n_4\\-n_4&n_3+n_4\end{pmatrix}^{-1}
\label{eq:GSM}
\end{equation}
The first decent approximant for the twist angles $(1.4^\circ,0,1.9^\circ)$ is given by $(n_1,n_2,n_3,n_4)=(0,3,0,-4)$, with $s=1.000456$ and $\delta\theta_2=-0.0995^\circ$, which we refer to as the 3-4 approximant.
Finally, in order to simplify our calculations, we rotate the entire system, $\theta_\ell\rightarrow\theta_\ell+\theta_\mathrm{SM}$, by $\theta_\mathrm{SM}=\frac{\pi}{2}-\atan (G_{\mathrm{SM},22}/G_{\mathrm{SM},12})$ such that, in the rotated problem, $\mathbf{G}_\mathrm{SM}=g_{\mathrm{SM}}\begin{pmatrix}\frac{\sqrt{3}}{2} & 0 \\ -\frac{1}{2} & 1\end{pmatrix}$ has a simple form.
Here, $g_{\mathrm{SM}}$ is the magnitude of the supermoir\'e reciprocal lattice vector, given by
\begin{equation}
g_{\mathrm{SM}}=g_0\sqrt{\frac{1+s^2-2s\cos(\theta_1-\theta_2)}{n_1^2+n_2^2+n_1 n_2}}
\end{equation}
where $g_0=4\pi/(\sqrt{3}a)$.
For future convenience, we further define $\Delta_x=\sqrt{3}g_{\mathrm{SM}}/2$ and $\Delta_y=g_{\mathrm{SM}}/2$.

We model the system at $B_\perp=0$ as an intralayer part,
\begin{equation}
H_{\ell}(k) = v_\mathrm{F} \sigma^+ e^{i\theta_\ell} [(k_x-K_{\ell,x}) - i(k_y-K_{\ell,y})] + h.c.
\end{equation}
where $\sigma^+ = \begin{pmatrix} 0&1\\0&0\end{pmatrix}$ is a raising operator acting on sublattice space, 
and ${K}_{\ell,\alpha} = (2{G}_{\ell,\alpha 1}+{G}_{\ell,\alpha 2})/3$.
The interlayer tunneling term from layer $\ell$ to $\ell^\prime$ is given by
\begin{equation}
H_{\ell^\prime \ell}(k) = w\sum_\nu \mathbf{T}_\nu e^{i \mathbf{q}_\nu^{\ell^\prime\ell}\cdot\mathbf{r}}
\end{equation}
where $\nu=1,2,3$ and $\mathbf{T}_\nu$ are matrices acting on sublattice space given by
\begin{eqnarray}
\mathbf{T}_1 = \begin{pmatrix}1&1\\1&1\end{pmatrix} ;
\mathbf{T}_2 = \begin{pmatrix}1&\omega^*\\\omega&1\end{pmatrix};
\mathbf{T}_3 = \begin{pmatrix}1&\omega\\\omega^*&1\end{pmatrix}
\end{eqnarray}
where $\omega=e^{2\pi i/3}$, and
\begin{equation}
(\mathbf{q}_1^{\ell^\prime\ell},
\mathbf{q}_2^{\ell^\prime\ell},
\mathbf{q}_3^{\ell^\prime\ell}) = \mathbf{G}_{\ell^\prime\ell}\begin{pmatrix}0&1&1\\0&0&1\end{pmatrix}
\end{equation}
Since $\mathbf{G}_{\ell^\prime \ell}$ are integer multiples of $\mathbf{G}_\mathrm{SM}$ (obtained by inverting Eq.~\ref{eq:GSM}), $\mathbf{q}^{\ell^\prime\ell}_\nu$ are also integer multiples of $\mathbf{G}_{\mathrm{SM}}$.
We have assumed a particular stacking configuration in which both moir\'e patterns have an $AA$ stacking region at the origin.
We choose $v_\mathrm{F}=\SI{1e6}{m/s}$ and $w=\SI{0.1}{eV}$ to match our previous calculations.

We apply a magnetic field via minimal substitution with a vector potential in the symmetric gauge: $k_x\rightarrow \pi_x=k_x+By/2$ and $k_y\rightarrow \pi_y=k_y-Bx/2$, satisfying $[\pi_x,\pi_y]=iB$.
We further define $X=k_x-By/2$ and $Y=k_y+Bx/2$ which commute with $\pi_{\alpha}$ and satisfy $[X,Y]=-iB$.
To simplify the following, we further define shifted operators $\pi_{\ell,x}=\pi_x-K_{\ell,x}$ and $\pi_{\ell,y}=\pi_y-K_{\ell,y}$.
These shifted operators are related to the unshifted ones by a unitary transformation, $\pi_{\ell,\alpha}=U_\ell \pi_\alpha U_\ell^\dagger$, where $U_\ell=e^{\frac{i}{B}[-K_{\ell,x}\pi_y+K_{\ell,y}\pi_x]}$.

The intralayer Hamiltonian becomes
\begin{equation}
\begin{split}
H_\ell &= v_\mathrm{F} e^{i\theta_\ell} \sigma^+ [\pi_{\ell,x} - i\pi_{\ell,y}] + h.c.
\\
&\equiv v_\mathrm{F}\sqrt{2B}e^{i\theta_\ell}\sigma^+ a^\dagger_\ell + h.c.
\end{split}
\end{equation}
where $a^\dagger_\ell = \frac{1}{\sqrt{2B}}[\pi_{\ell,x} - i\pi_{\ell,y}]$.

The interlayer tunneling consists of terms $e^{i\mathbf{q}\cdot\mathbf{r}}$, which can each be expanded
\begin{equation}
\begin{split}
 e^{i\mathbf{q}\cdot\mathbf{r}}=
 e^{\frac{i}{B}[{q}_{x}Y - q_{y}X]} 
 e^{\frac{i}{B}[-{q}_{x}\pi_y + q_{y}\pi_x]} 
\end{split}
\end{equation}
into a product of two factors acting on the Hilbert spaces of $(X,Y)$ and $(\pi_x,\pi_y)$ separately.

Working in the eigenbasis of $X$, we have for the first factor
\begin{equation}
 \bra{X+q_x}e^{\frac{i}{B}[{q}_{x}Y - q_{y}X]} \ket{X} = 
 e^{\frac{i}{B}[-q_{y}(X+q_x/2)]} 
\end{equation}
and all other matrix elements zero.  
Since $\mathbf{q}$'s appearing in the Hamiltonian are of the form $\mathbf{q}^{\ell^\prime\ell}_{\nu}=(d^{\ell^\prime\ell}_{\nu,x} \Delta_x, d^{\ell^\prime \ell}_{\nu,y} \Delta_y)$, where $d^{\ell^\prime\ell}_{\nu,\alpha}$ are integers,
 the interlayer Hamiltonian only connects states $\ket{X}$ with all other states $\ket{X+j \Delta_x}$.
For particular magnetic fields $B_\perp=p\Delta_x\Delta_y /(2\pi q)$, where $p,q$ are integers, the Hamiltonian becomes symmetric with respect to a shift $\ket{X}\rightarrow \ket{X+p \Delta_x}$.
This allows us to Fourier transform along the $X$ direction, labeling states as
$\ket{j;x_0k_0} \propto \sum_{m} e^{\frac{2\pi i k_0}{p}(x_0+j+mp)}\ket{X=(x_0+j+mp)\Delta_x}$,
where the offset $x_0\in[0,1]$ and momentum $k_0\in[0,1]$ are good quantum numbers of $H$.

For the second factor, we work in the layer-dependent eigenbasis of $n_\ell=a^\dagger_\ell a_\ell$ and, since this term tunnels between layers, we are concerned with the matrix element 
\begin{equation}
\begin{split}
 \bra{n_{\ell^\prime}^\prime}&e^{\frac{i}{B}[-{q}_{x}\pi_y + q_{y}\pi_x]} \ket{n_\ell} \\
 &=\bra{n^\prime}U_{\ell^\prime}^\dagger e^{\frac{i}{B}[-{q}_{x}\pi_y + q_{y}\pi_x]} U_\ell \ket{n} \\
 &= 
 \bra{n^\prime} e^{\frac{i}{B}[-({q}_{x}+K_{\ell,x}-K_{\ell^\prime,x})\pi_y + (q_{y}+K_{\ell,y}-K_{\ell^\prime,y})\pi_x]}  \ket{n} \\
 &\equiv D_{n^\prime n}\left(\frac{[q_x+K_{\ell,x}-K_{\ell^\prime,x} + i(q_y+K_{\ell,y}-K_{\ell^\prime,y})]}{\sqrt{2B}}\right)
 \end{split}
\end{equation}
where 
\begin{equation}
\begin{split}
D_{n^\prime n}(z) &= \bra{n^\prime} e^{z a^\dagger - z^* a} \ket{n} \\
&= \begin{cases}
e^{-\frac{|z|^2}{2}} (z)^{n^\prime-n} \sqrt{\frac{n^\prime!}{n!}}L^{(n^\prime-n)}_n(|z|^2) & \mathrm{if}\; n^\prime\geq n\\
e^{-\frac{|z|^2}{2}} (-z^*)^{n-n^\prime} \sqrt{\frac{n!}{n^\prime!}}L^{(n-n^\prime)}_{n^\prime}(|z|^2) & \mathrm{else}
\end{cases}
\end{split}
\end{equation}
and $L_n^{(\alpha)}(x)$ are the generalized Laguerre polynomials.

Putting everything together, we choose as our full computational basis the set of states $\ket{n_\ell,j,\sigma,\ell;x_0k_0}$,
and diagonalize the Hamiltonian sampling over the space of $x_0,k_0\in[0,1]$.
The Hamiltonian is $H=\sum_\ell H_\ell + \sum_{\langle\ell^\prime,\ell\rangle}H_{\ell^\prime \ell}$, where the sum over $\langle\ell^\prime,\ell\rangle$ is over adjacent layers.
Denoting the matrix elements of $H$ as
\begin{equation}
(H^{(x_0 k_0)})^{n^\prime j^\prime \sigma^\prime \ell^\prime}_{n j \sigma\ell}
\equiv \bra{n_\ell^\prime,j^\prime,\sigma^\prime,\ell^\prime;x_0,k_0}H
\ket{n_\ell,j,\sigma,\ell;x_0,k_0},
\end{equation}
the non-zero elements of the intralayer Hamiltonian are
\begin{equation}
\begin{split}
(H^{(x_0k_0)})^{n+1,j,2,\ell}_{n,j,1,\ell}
&=v_\mathrm{F}\sqrt{2B(n+1)}e^{i\theta_\ell}\\
&=[(H^{(x_0k_0)})^{n,j,1,\ell}_{n+1,j,2,\ell}]^*
\end{split}
\end{equation}
and the interlayer matrix elements are, for $|\ell^\prime-\ell|=1$,
\begin{eqnarray}
&(H^{(x_0k_0)}&)^{n^\prime j^\prime\sigma^\prime\ell^\prime}_{nj\sigma\ell}\\
&=
w\sum_{\nu} &(\mathbf{T}_{\nu})_{\sigma^\prime\sigma} 
\delta_{j^\prime,j+d_x}
\nonumber
\\
&&\times e^{-\frac{2\pi i q}{p} d_y(x_0+j+\frac{d_x}{2})-\frac{2\pi i k_0 d_x}{p}} D_{n^\prime n}(z)
\nonumber
\end{eqnarray}
where we have used the shortened notation $d_\alpha=d^{\ell^\prime \ell}_{\nu,\alpha}$ and
\begin{equation}
z=\frac{1}{\sqrt{2B}}[\Delta_x d_x+K_{\ell,x}-K_{\ell^\prime,x} + i(\Delta_y d_y+K_{\ell,y}-K_{\ell^\prime,y})]
\end{equation}

\begin{figure*}
    \centering
    \includegraphics[width=1\linewidth]{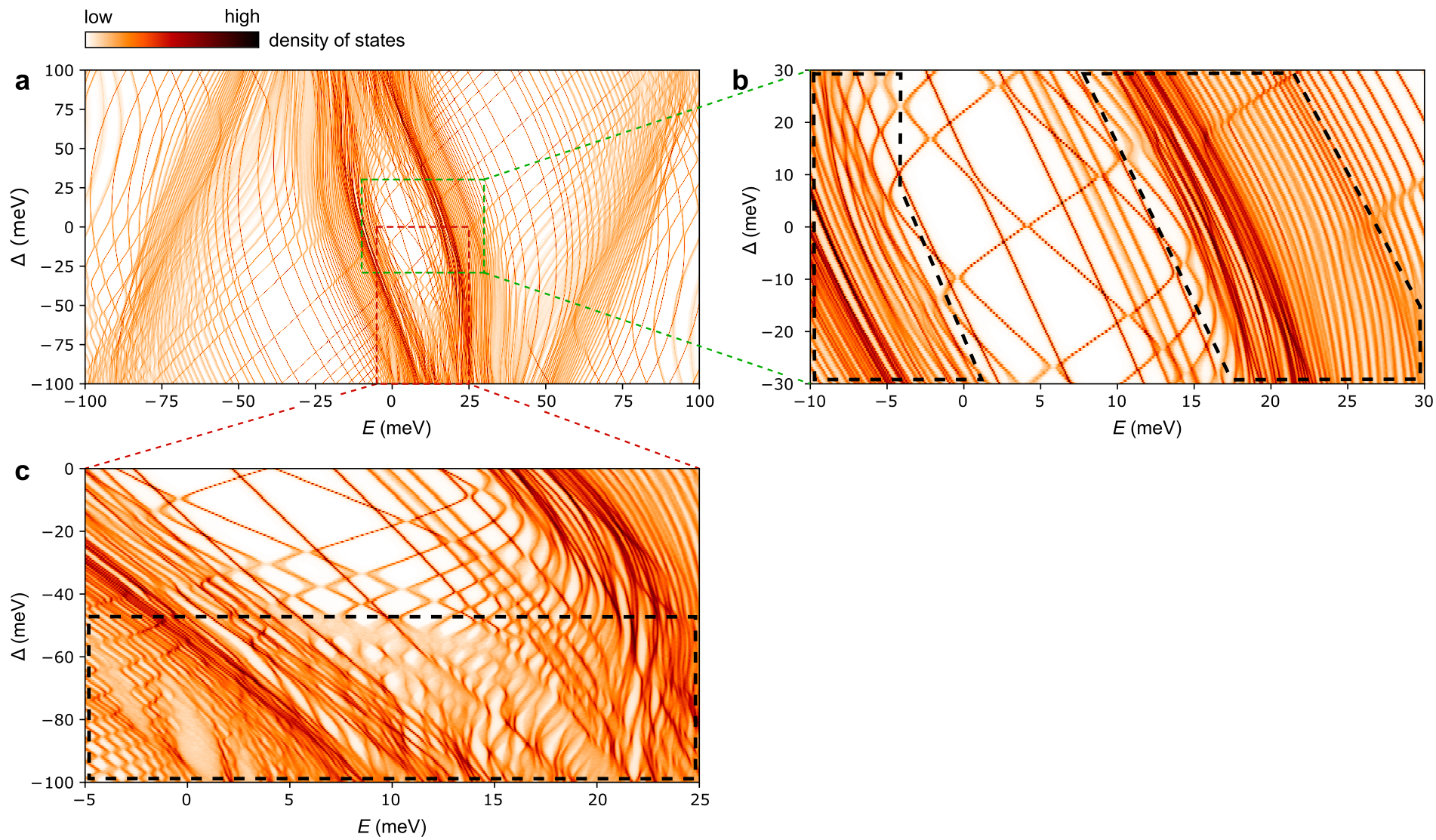}
    \caption{
    \figtitle{Density of states under finite magnetic field.}
    \panel{a} DOS calculation for a 3-4 approximant (Methods~\ref{ssec:SF_finite_B}) under $B_\perp=\SI{1}{T}$.
    \panel{b} Same as \panel{a}, zooming in on moderate $\Delta$. Black dashed lines approximately enclose the electronic quasiperiodic regimes as a guide to the eye.
    \panel{c} Same as \panel{a}, zooming in on $\Delta<0$ and low energies.
    }
    \label{fig:SF-finite-B}
\end{figure*}

For the 3-4 approximant, we have that the accessible magnetic fields are of the form $B\approx(\SI{3.01}{T})(p/q)$.
The results shown in Fig~\ref{fig:SF-finite-B} were performed at $p=1,q=3$, which correspond to $B\approx \SI{1}{T}$.  
The LL Hilbert space is truncated at $N_{\mathrm{max}}=1000$.  
We additionally add a layer potential $(-\Delta,0,\Delta)$ for the three layers to model the displacement field.
For each $x_0,k_0$, we obtain a set of eigenvalues $\{E_i^{x_0k_0}\}$.
The density of states is then obtained by averaging over $x_0$ and $k_0$,
\begin{equation}
\mathrm{DOS}(E)=\frac{4}{2q A_\mathrm{uc}}\left\langle \sum_i \delta(E-E^{x_0 k_0}_i)\right\rangle_{x_0 k_0}
\end{equation}
where $A_{\mathrm{uc}}=8\pi^2/(\sqrt{3} g_{\mathrm{SM}}^2)$ is the super-moir\'e unit cell area and the factor of $4$ accounts for spin-valley degeneracy.

\extlbl~\ref{fig:SF-finite-B} shows the calculated DOS vs energy and layer potential, $\Delta$.
At moderate displacement field, $|\Delta| \lesssim \SI{30}{meV}$ (\extlbl~\ref{fig:SF-finite-B}\panel{b}) we find a transition between two regimes.
At low energies, $\SI{-3}{meV} \lesssim E \lesssim \SI{12}{meV}$, the periodic-like regime gives rise to well-defined LLs.
Outside this energy window the quasiperiodicy is strongly pronounced and results in broadened LLs (black dashed boxes in \extlbl~\ref{fig:SF-finite-B}\panel{b}).
Another quasiperiodic regime is found at high layer potential assymetry, $|\Delta| \gtrsim \SI{50}{meV}$, $|D/\epsilon_0| \gtrsim \SI{0.7}{V/nm}$ (Fig.~\ref{fig:LLs}\panel{a}, Fig.~\ref{fig:bands}\panel{c}) and lower carrier densities and energies, $n_\text{tot} \lesssim n_\text{s,12}$, $|E| \lesssim \SI{25}{meV}$ (Fig.~\ref{fig:bands}\panel{c}).
At finite magnetic field we again find quasiperiodic HSBs grouped into broadened LLs (\extlbl~\ref{fig:SF-finite-B}\panel{c}, dashed yellow).

\section*{Acknowledgements}
We thank Benjamin E. Feldman, Mikito Koshino, Ziyan Zhu, D. Kwabena Bediako, Allan H. MacDonald, Francisco Guinea, Leonid Levitov, Leonid Glazman, and Erez Berg for illuminating discussions.
AU acknowledges support from the MIT Pappalardo Fellowship and from the VATAT Outstanding Postdoctoral Fellowship in Quantum Science and Technology.
MTR acknowledges support from the MIT Pappalardo Fellowship.
This work was supported by the Army Research Office MURI W911NF2120147 (AU), the National Science Foundation (DMR-1809802, MR and DRL), the STC Center for Integrated Quantum Materials (NSF grant no. DMR-1231319, SCdlB, TD, PJDC, and NP), and the Gordon and Betty Moore Foundation’s EPiQS Initiative through grant GBMF9463 to PJH.
This work was performed in part at the Harvard University Center for Nanoscale Systems (CNS), a member of the National Nanotechnology Coordinated Infrastructure Network (NNCI), which is supported by the NSF under NSF ECCS award no. 1541959.
KW and TT acknowledge support from JSPS KAKENHI (Grant Numbers 19H05790, 20H00354 and 21H05233).
RL is supported by the Israel Science Foundation (ISF) through Grant No. 1259/22.

\setcitestyle{numbers}
\bibliography{references}

\clearpage

\clearpage
\SupplementaryMaterials
\section*{Supplementary Information}

\subsection{Spectral function constant-energy slices}\label{ssec:SF-2D-cuts}
\supplbl~\ref{fig:SF-2D-cuts} shows constant-energy slices of the SF versus $(k_x,k_y)$.
Closed Fermi surfaces are visible at low energies, $|\varepsilon| \lesssim \SI{4}{meV}$, whereas at higher energies open Fermi surfaces dominate. 
This accounts for the well-defined LLs observed at low densities in our magnetotransport (Fig.~\ref{fig:LLs}\panel{a}).
The Fermi surface of layer~2 (green) breaks up at a lower energy compared with layers 1 and 3, accounting for the layer~2 LLs disappearance at density $n_\text{tot}$ lower than for layer 1 and 3 LLs (Fig.~\ref{fig:LLs}\panel{a-c}).
See Supplementary Video 2 for the full sequence.

\subsection{Other twist angles in quasiperiodic regime}\label{ssec:other-angles}
\begin{figure*}
    \centering
    \includegraphics[width=180mm]{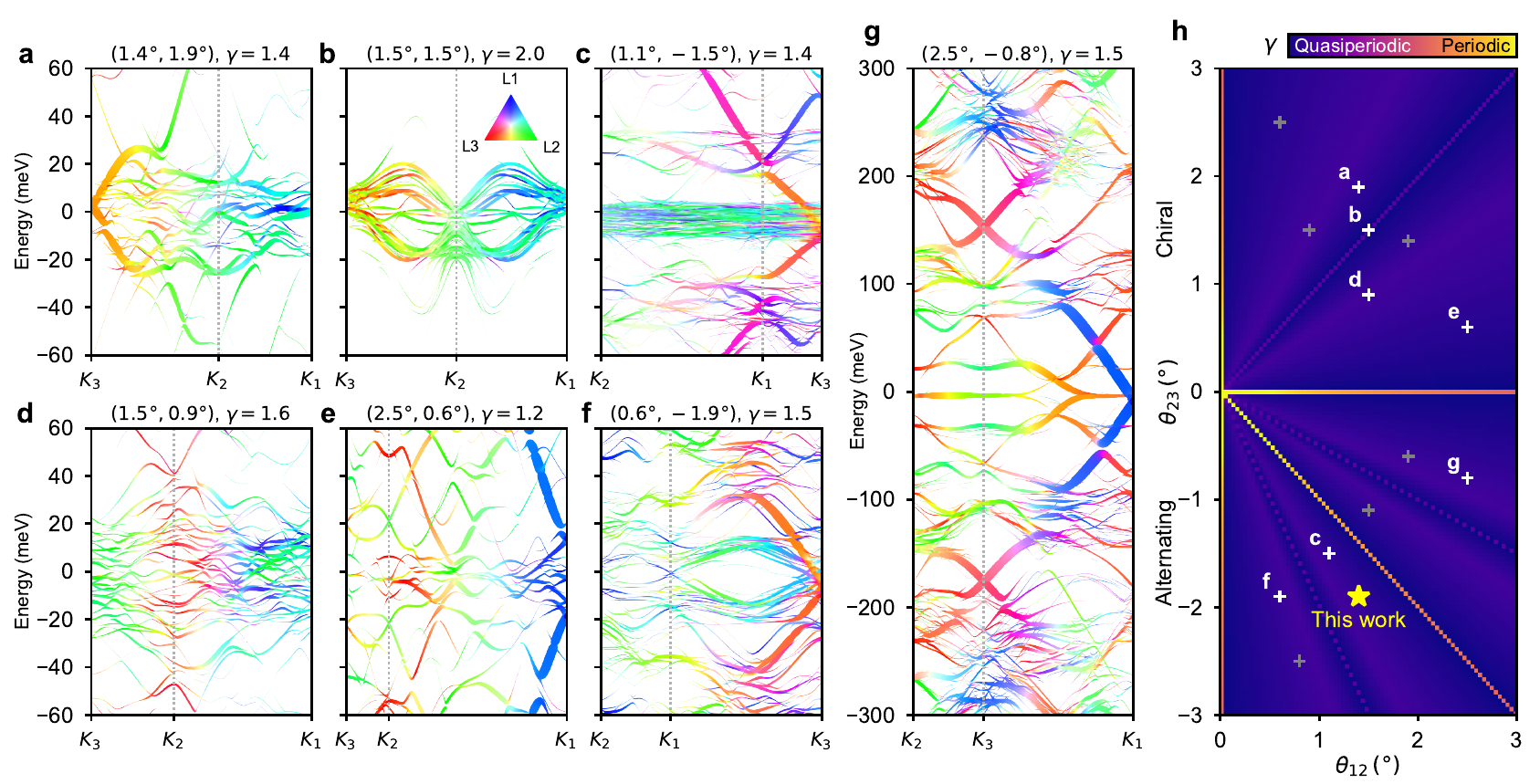}
    \caption{
    \figtitle{Spectral function calculations for other angles.}
    \panel{a-g} SF calculated for twist angles $(\theta_{12},\theta_{23})$ as indicated.
    \panel{a,b,d,e} chiral setups, $\operatorname{sgn}(\theta_{12})=\operatorname{sgn}(\theta_{23})$, and \panel{c,f,g} alternating setups, $\operatorname{sgn}(\theta_{12})\neq\operatorname{sgn}(\theta_{23})$. Shown are momentum-space traces between the three $K_i$ points of layers $i = \lbrace1,2,3\rbrace$.
    \panel{h} Separation of scales, $\gamma$, versus $\theta_{12}$ and $\theta_{23}$. White crosses indicate the angles considered in panels \panel{a-g}.
    Gray crosses are twin angles describing the same systems as the white crosses ($\theta_{12} \longleftrightarrow \theta_{23}$ for chiral setups and $\theta_{12} \longleftrightarrow -\theta_{23}$ for alternating setups).
    }
    \label{fig:other-angles}
\end{figure*}

We perform SF calculations for selected angle pairs $(\theta_{12},\theta_{23})$ (\supplbl~\ref{fig:other-angles}).
Importantly, all selected angles show low separation of scales, $1<\gamma\leq 2$, simultaneously with a high degree of quasiperiodicity, expressed by the multiple gaps and thin line widths (small point sizes) in the calculated SF.
This supports the use of $\gamma$ as a proxy for \moire quasiperiodicity.
The calculations were performed as described in Methods~\ref{ssec:SF}, using the same microscopic parameters.
Note that these parameters may not reliably predict the bandwidths due to different Fermi velocity renormalizations induced by self-screening.

\subsection{Evidence against full lattice relaxation}\label{ssec:relaxation}
Our separation of length scales argument as well as our SF calculations do not include the effects of lattice relaxation \cite{zhu2020modeling} which is expected to be pronounced at small twist angles, $|\theta_{ij}|\lesssim \SI{1}{\degree}$.
Specifically, recent experiments show \cite{turkel2022orderly} that in nearly symmetric magic-angle TTG lattice relaxation forms domains of mirror-symmetric TTG, bounded by domain walls of varying twist angles.
These domains of ``full relaxation'' seem to occur at very small twist angles between the outer layers, $|\theta_{13}|\lesssim \SI{0.3}{\degree}$.
Deviating further from the mirror-symmetric configuration, the domains are expected to shrink, keeping an approximately constant domain wall width \cite{zhu2020modeling}, until eventually the symmetric configuration disappears everywhere.
In our case, since $\theta_{13}=\SI{0.5}{\degree}$ is quite large, we do not expect full lattice relaxations.
Indeed, our measurements clearly show: (i) separate sets of resistive peaks that correspond to $\theta_{12}$ and $\theta_{23}$, (ii) three independent Dirac cones, and (iii) features of quasicrystallinity in the LLs.
This is in sharp contrast to transport measurements on nearly mirror-symmetric TTG \cite{park2021tunable,hao2021electricfield}, suggesting that our structure does not form domains of fully relaxed mirror-symmetric TTG.
The electronic behavior is therefore expected to transition from periodic-like to \moire quasicrystalline at a small angle $\SI{0.3}{\degree} \lesssim |\theta_{13}| \lesssim \SI{0.5}{\degree}$ \cite{hao2021electricfield,turkel2022orderly}.
It is possible that other two-angle TTG systems will also exhibit relaxation into domains of higher-order periodic approximants (not necessarily mirror-symmetric TTG) when such relaxation involves small deformations and low elastic energy costs.
A better estimation of the transition angle as well as other effects of relaxation on the electronic properties are outside the scope of this work.

\subsection{Supplementary Video 1}\label{ssec:movie-1}
Supplementary Video 1 shows the SF calculated for $(\theta_{12},\theta_{23})=(\SI{1.4}{\degree},\SI{-1.9}{\degree})$ along a line cut through $K_3-K_1-K_2$ for varying values of the layer potential $\Delta$, between \SI{-100}{meV} and \SI{100}{meV}.
It demonstrates the high degree of tunability of the \moire quasicrystal accessible with electrostatic gating.

\subsection{Supplementary Video 2}\label{ssec:movie-2}
\begin{figure*}
    \centering
    \includegraphics[width=1\linewidth]{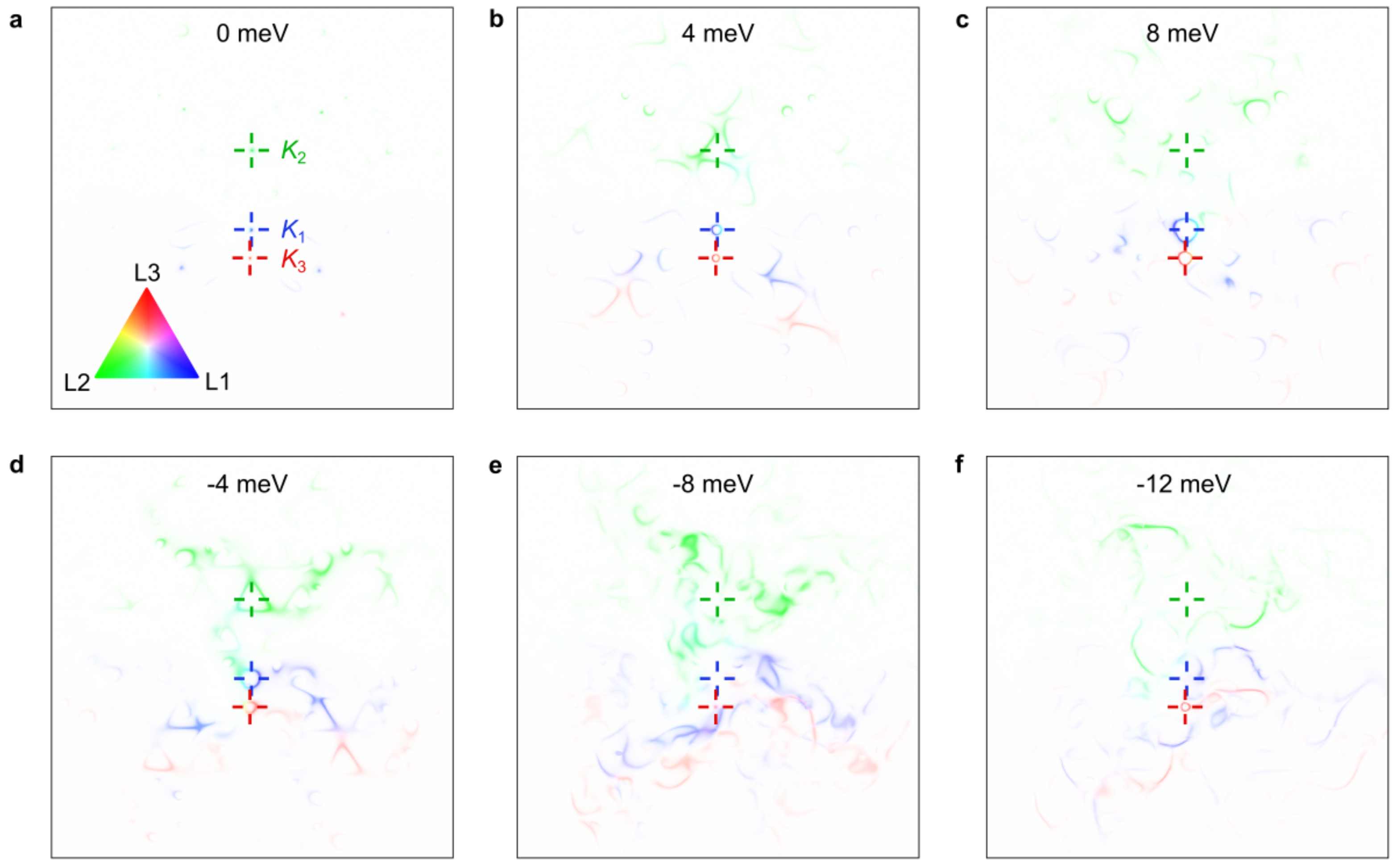}
    \caption{
    \figtitle{SF equi-energy cuts for $\Delta=0$.}
    Spectral function versus $k_x$ and $k_y$ calculated for $(\theta_{12},\theta_{23}) = (\SI{1.4}{\degree},\SI{-1.9}{\degree})$, showing constant energy cuts at the indicated energies.
    Color describes the relative weight of the SF on the different layers according to the colored triangle in \panel{a}, inset (see Methods~\ref{ssec:SF}).
     }
    \label{fig:SF-2D-cuts}
\end{figure*}

Supplementary Video 2 shows constant energy cuts of the SF calculated for $(\theta_{12},\theta_{23})=(\SI{1.4}{\degree},\SI{-1.9}{\degree})$ for energies between \SI{-30}{meV} and \SI{30}{meV} (see also \supplbl~\ref{fig:SF-2D-cuts}).

\end{document}